\documentclass{article}
\usepackage[utf8]{inputenc}
\usepackage{float}
\usepackage{amsmath}
\usepackage{chngcntr}
\usepackage{tabularx}
\usepackage{xcolor}
\usepackage{caption}
\usepackage{subcaption}
\usepackage{svg}
\usepackage{mathtools}
\usepackage{hyperref}
\hypersetup{
    colorlinks=true,
    linkcolor=blue,
    filecolor=magenta,      
    urlcolor=cyan,
    pdftitle={Overleaf Example},
    pdfpagemode=FullScreen,
}

\renewcommand{\figurename}{\bf Figure}
\renewcommand{\tablename}{\bf Table}

\title{Measuring Network Dynamics of Opioid Overdose Deaths in the United States}
\author{\normalsize{Kushagra Tiwari$^{1\ast}$, M. Amin Rahimian$^{1\ast}$, Mark S. Roberts$^{2}$}, \\ \normalsize{Praveen Kumar$^{2}$ and Jeannine M. Buchanich$^{3}$}\\
\footnotesize{$^1$Department of Industrial Engineering, $^2$Department of  Health Policy and Management,}\\ \footnotesize{and $^3$Department of  Biostatistics \vspace{-1mm}}\\ \footnotesize{ University of Pittsburgh \vspace{-1mm}}\\
\footnotesize{$^\ast$~To whom correspondence should be addressed; emails: kut20@pitt.edu, rahimian@pitt.edu.}}
\date{}

\usepackage{graphicx}
\usepackage{amsmath}

\begin{document}

\maketitle
\vspace{-0.3in}
\begin{abstract}
The US opioid overdose epidemic has been a major public health concern in recent decades. There has been increasing recognition that its etiology is rooted in part in the social contexts that mediate substance use and access; however, reliable statistical measures of social influence are lacking in the literature. We use Facebook's social connectedness index (SCI) as a proxy for real-life social networks across diverse spatial regions that help quantify social connectivity across different spatial units. This is a measure of the relative probability of connections between localities that offers a unique lens to understand the effects of social networks on health outcomes. We use SCI to develop a variable, called ``deaths in social proximity'', to measure the influence of social networks on opioid overdose deaths (OODs) in US counties. Our results show a statistically significant effect size for deaths in social proximity on OODs in counties in the United States, controlling for spatial proximity, as well as demographic and clinical covariates. The effect size of standardized deaths in social proximity in our cluster-robust linear regression model indicates that a one-standard-deviation increase, equal to $11.70$ more deaths per $100,000$ population in the social proximity of ego counties in the contiguous United States, is associated with thirteen more deaths per $100,000$ population in ego counties. To further validate our findings, we performed a series of robustness checks using a network autocorrelation model to account for social network effects, a spatial autocorrelation model to capture spatial dependencies, and a two-way fixed-effect model to control for unobserved spatial and time-invariant characteristics. These checks consistently provide statistically robust evidence of positive social influence on OODs in US counties. Our analysis provides a pathway for public health interventions informed by social network structures. The statistical robustness of our primary variable of interest, deaths in social proximity, supports the hypothesis of a social network effect on OODs. Using agent-based modeling (ABM) to simulate social networks can offer an effective method to design interventions that incorporate the dynamics of social networks for maximum impact.


\end{abstract}
\section*{Introduction}
The opioid overdose epidemic is a major public health crisis in the US, with an exponentially increasing number of drug overdose deaths in the last four decades \cite{jalal2018changing,jalal2022exponential}. Alpert et al. (2022) report that opioid overdose deaths (OODs) account for 75\% of the increase in drug overdose deaths \cite{alpert2022origins}. Addressing this crisis requires planned interventions that focus on supply-side regulations and the dynamics of social connections. The rate of initiation of opioid misuse is known to increase due to social influence \cite{lim2022modeling}. Costello et al. (2021) report that of the 370 participants who entered an opioid withdrawal program, $97\%$ identified knowing the individual with whom they initiated opioid use, with friendship being the most reported relationship between participants and their initiation partners \cite{costello2021peer}. Similarly, Rigg et al. (2018) note that two-thirds of misused drugs are obtained from friends and family \cite{rigg2018opioid}. Guarino et al. (2018) study of $539$ young adults who misuse opioids and heroin indicates that their first experiences with the misuse of prescription opioids typically occur in a social setting with peers \cite{guarino2018young}. The misuse of prescription opioids has been growing among young people \cite{jalal2020age}. Syvertsen et al. (2017) make similar observations about young people who experiment with drugs and the initiation of drug use \cite{syvertsen2017down}. Social networks can have positive and negative impacts on substance use. Empirical results have indicated that peer networks with subjects who do not use substances have a positive influence on curbing drug abuse; however, networks consisting of more substance users are likely to increase substance use \cite{valente2007peer}. In intervention designs, recovery support strategies, including peer recovery, have shown encouraging results. Peer workers who have completed their recovery help others in recovery from substance addiction. This form of peer-supported recovery is found to be more effective in reducing the prevalence of opioid use disorder (OUD) \cite{stringfellow2022reducing}. 

Providing reliable measures of social influence on the opioid epidemic is complicated by the confounding factors that influence opioid misuse and social interactions, on the one hand, and the ethical barriers to randomized experiments, on the other. Studies measuring social influence use complex systems-based generative models to understand these phenomena in areas such as voting contagion \cite{braha2017voting}. Specifically, Braha and Aguiar (2017) \cite{braha2017voting} use a generalized social voter model combined with spatial-statistical analysis to examine how social influence has shaped voting behavior in the US presidential election over the past century. They distinguish between social contagion and external influences (e.g., media and opinion leaders) to assess their impact on county-level vote share distributions over time and geography. By analyzing spatial patterns, they demonstrate that social influence is geographically clustered and spreads like a contagion across county borders. In contrast, applying similar modeling techniques to the opioid epidemic to design targeted interventions has encountered certain limitations. Homer et al. (2021) \cite{homer2021dynamic} discuss the complexity of modeling OUD and highlight the limitations of their models in accurately capturing real-life scenarios due to their simplicity. Recognizing the complexities of modeling the opioid epidemic through generative processes, our research aims to address these gaps using controlled regressions. 

From the traditional setting to the new digital era of social networks, we have witnessed a significant shift in informal social interactions. Facebook is the largest informal online social network globally, with $2.1$ billion active users and $239$ million active users in the US and Canada. Given its broad reach and prevalence, Facebook connections can provide insight into real-life social networks in many geographical regions. Facebook has released a social connectedness index dataset that measures how intensely geographical locations are connected according to the relative probability of connections \cite{bailey2018social}. In Supplementary Information section \ref{sec:lirev} we provide a review of SCI use cases as a proxy for real-life social connections in public health, epidemiology, economics and development applications. In the following paragraph, we justify our use of SCI for measuring the network dynamics of OODs in the US.

We use SCI to construct a measure of OODs in the social proximity of counties in the United States and investigate the statistical significance of its effect on county-level OODs after controlling for clinical, spatiotemporal, and socioeconomic confounders. Our analysis seeks to provide information on the role of social influence on OODs. Using SCI to represent the intensity of interpersonal networks between counties reflects the possibility of physical interactions between county residents. Bailey et al. (2020) \cite{bailey2020social,bailey2020determinants} show that SCI is predictive of travel patterns within urban areas and throughout Europe. Coven et al. (2023) \cite{coven2023jue} show that counties with higher social connections to New York City are the most preferred destinations for those moving away from the city during the pandemic, highlighting the association of physical interaction with SCI. Kuchler et al. (2022) \cite{kuchler2022jue} use SCI to show that COVID-19 is more likely to spread between regions with stronger social networks and highlighted the potential of SCI to improve the prediction of epidemics. Although SCI provides a robust measure of social connectedness, some studies have explored alternative data sources, such as social media platforms, to examine trends in the context of the opioid epidemic. However, these approaches have faced significant challenges due to demographics and other data limitations \cite{ge2024reddit, pandrekar2018social, balsamo2021patterns}. For example, Pandrekar et al. (2018) \cite{pandrekar2018social} use Reddit data to analyze psychological categories and patterns of opioid abuse on a national scale. A major limitation of their study is that the data collected through the Reddit API do not provide access to user location information. In addition, the Reddit data do not indicate whether users are friends, which restricts the ability to analyze the structure of social networks and their association with OODs.

In contrast, SCI offers distinct advantages. SCI is location-based and provides detailed information about the structure of social networks in US counties. SCI measures friendship networks, serving as a proxy for real-world social connections. Unlike Reddit data, SCI allows for the analysis of location-specific friendship networks, which makes it particularly useful for studying how social networks influence the opioid epidemic on a population scale. Our choice of SCI as the network measure is informed by previous use cases that reflect the real-life dynamics of social connections in different domains such as education \cite{diemer2022no} and public health \cite{kuchler2022jue}. Our objective is to measure how these social connections contribute to heterogeneous patterns of opioid overdose deaths in US counties.

In our study, we present a novel perspective on analyzing the association between friendship networks and opioid overdose deaths at the population level in counties within three distinct geographical regions: the eastern United States, the central and western United States, and the entire contiguous United States. We provide statistical evidence linking the geographical spread of OODs with the structure of social connections.

To achieve this, we use a range of statistical methodologies. We use cluster-robust linear regression to account for intra-cluster correlation between counties within the same states, network and spatial autocorrelation methods to address the autocorrelation in error terms arising from unobserved factors shared among spatially or socially connected units, and two-way fixed effects models to control for unobserved spatial and temporal heterogeneities. Additionally, we perform robustness checks to account for the distance decay of proximity weights and apply a two-stage least squares method to jointly address spatial and network effects, discussed in the Supplementary Information sections \ref{sec:distance decay weight check} and \ref{sec:G2sls}. Our multifaceted statistical analysis demonstrates that our variable of interest, deaths in social proximity, is statistically significant across the three distinct geographical regions.

Understanding the role of social networks is important in designing interventions to combat opioid misuse behavior. Research measuring the network dynamics of opioid overdose death on the US population scale and at the resolution of counties remains limited, with the exception of \cite{mackle2022social,cutler2024thick} that measure social network drivers of the opioid epidemic and use natural experiments to support their claims. M{\"a}ckle and Reunzi (2023) examine changes in county-level overdose deaths due to the reformulation of OxyContin and the must-access Prescription Drug Monitoring Program (PDMP) \cite{mackle2022social}. They analyze policy-induced shocks to estimate the indirect effect of friendship networks (measured by SCI) on OODs. Their analysis includes the correlation of ``social proximity" with OODs by using a two-way fixed-effects model. Compared to M{\"a}ckle and Reunzi (2023) \cite{mackle2022social}, our study differs in the methodological approach. Along with cluster-robust linear regression and two-way fixed effects, our analysis uses a network and spatial autocorrelation model to account for autocorrelation in error terms, which M{\"a}ckle and Reunzi (2023) \cite{mackle2022social} do not consider when investigating the association of ``social proximity" with OODs. We also include a robustness check to account for spatial and network autocorrelation together and provide evidence that the statistical significance of the ``deaths in social proximity" variable is not sensitive to the choice of distance decay function.
Furthermore, our regression includes domain-specific covariates such as availability of naloxone and buprenorphine, opioid dispensing rate, mental health distress rate and state-level fentanyl and analog seizure data to account for illicit opioids which M{\"a}ckle and Reunzi (2023)  \cite{mackle2022social} do not consider when investigating the association between ``social proximity" and OOD. Furthermore, M{\"a}ckle and Reunzi (2023) use the National Vital Statistics System (NVSS) public database and employ a backout procedure to recover mortality data points that the NVSS otherwise suppresses if the county has fewer than ten deaths. In contrast, our study uses mortality data from the National Center for Health Statistics (NCHS) which provides access to these suppressed data points. In this regard, our results complement their findings, as their coefficient for ``social proximity" is also statistically significant. Concurrently with our work, Cuttler and Donahoe (2024) \cite{cutler2024thick} have explored the dynamics of opioid death rates, focusing on SCI and the distance between counties to analyze spillovers. They posit that the increase in opioid demand is endogenous, resulting from spillovers between affected populations. Their study underscores the importance of social and spatial spillovers in estimating opioid demand, which is correlated with increased mortality rates. However, our work differs in its methodological approach, concentrating on constructing a socially lagged variable to assess the impact of social influence on overdose death rates. 

In the following section, we formally introduce our variable of interest, ``deaths in social proximity''. The descriptive statistics serve as a starting point for later estimating the effect size of deaths in social proximity in the Results section. 

\section*{OOD rates in social and spatial proximity}
The root of the opioid epidemic is partially associated with social contexts that mediate substance use and accessibility. Existing studies integrating social network analysis with the geographical spread of overdose deaths have demonstrated how social characteristics influence the trajectory of substance use, for example, geographic discordance, which means that the community in which the overdose death occurs is different from the community of residence \cite{forati2023journey}. Using data on overdose deaths from 2017 to 2020 in Milwaukee County, Wisconsin, Forati et al. (2023) build a social-spatial network framework to detect network interaction hotspots for overdose deaths and analyze their geographical discordance \cite{forati2023journey}. However, their study is limited to Milwaukee County and does not extend to US population scales.

As discussed in the Introduction, the significant contribution of social influence in initiating substance use is well documented in the literature. Researchers have highlighted the impact of social networks on the regulation of substance use patterns based on the ego network \cite{koram2011role}. However, applying these findings to a geographical context presents challenges, as geographical proximity substantially influences social connections and communication patterns across varying distances \cite{xu2022beyond}. In our model, we control for the inherent spatial patterns of geographical proximity to refine our estimate of the effect of social networks on OODs. 

Assessing the strength of social ties within every individual's network in a wide geographical area is very resource-intensive. To address this challenge, one potential approach is to aggregate and estimate the social networks of metapopulations residing in different localities, such as ZIP codes or counties in the United States. This offers a broader perspective and alleviates some of the constraints associated with individual-level analysis. In 2018, Meta Platforms, which operates the Facebook social network, released a data set that measures the distribution of social ties of location-specific networks globally. SCI is a surrogate for real-life friendship networks between registered Facebook users at each location \cite{bailey2018social}. It quantifies the strength of friendship ties in various locations using relative probability and is available at the ZIP code and county level for the United States. Formally, the SCI between two locations $i$ and $j$ is defined as follows \cite{bailey2018social}:
\begin{align*}
\mbox{SCI}_{ij} &=\frac{\mbox{Facebook Connections}_{ij}}{\mbox{Facebook Users}_i \times \mbox{Facebook Users}_j}.
\end{align*}
Here, $\mbox{Facebook Users}_i$ represents the number of Facebook users in the county $i$. $\mbox{Facebook Connections}_{ij}$ is the total number of Facebook friendship connections between individuals in counties $i$ and $j$.

\begin{figure}[!hbt]
   \centering
   \includegraphics[width=\textwidth]{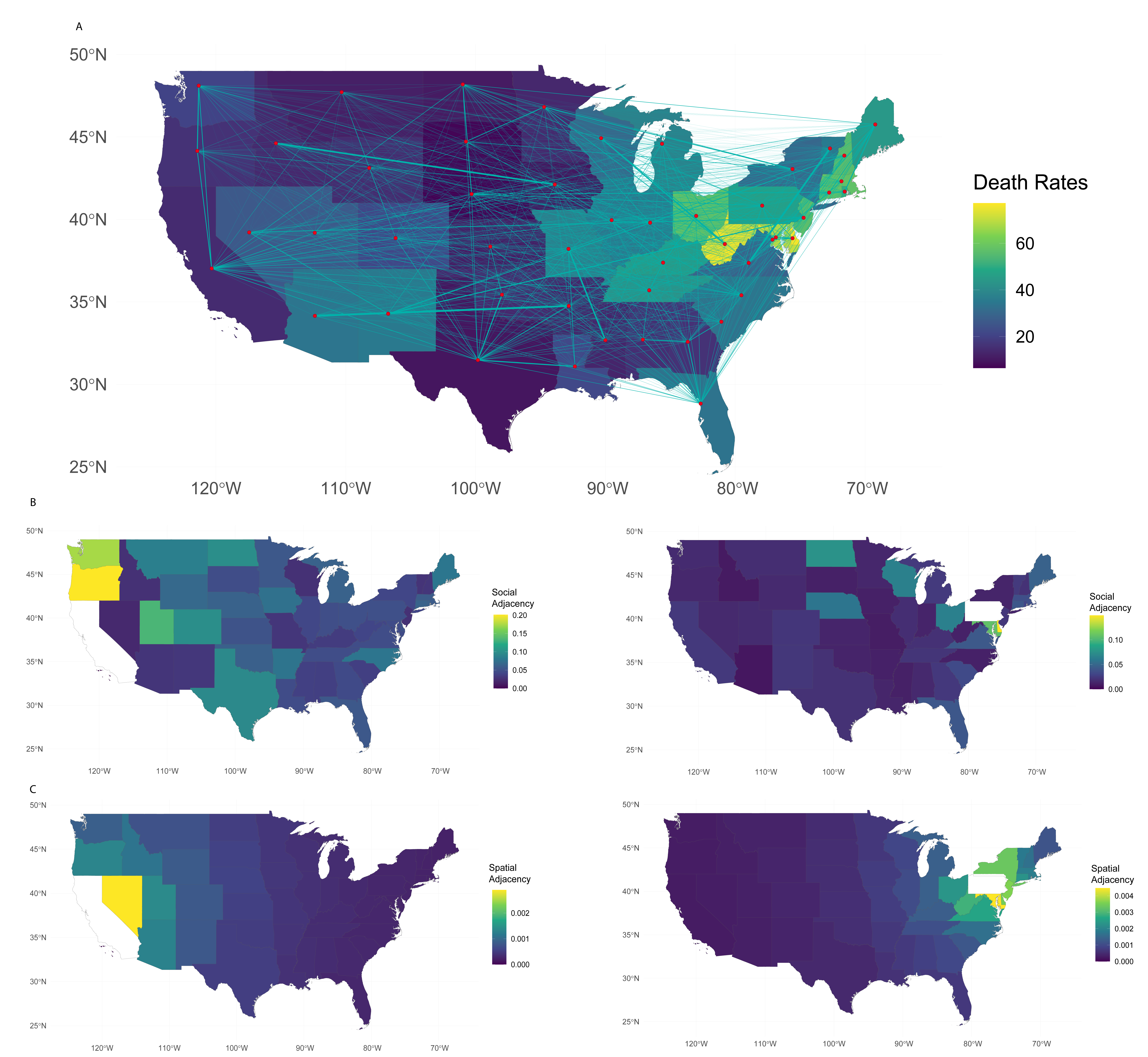}
\caption{
A) The spatial distribution of overdose death rates per 100,000 population in contiguous US from 2018 to 2019. Superimposed on this map is a social network diagram with edge widths representing the state-level social proximity weights . B) The two middle maps show the social proximity weights of alter states to California (on the left) and Pennsylvania (on the right). C) The bottom two maps show the spatial proximity weights of alter states to California (on the left) and Pennsylvania (on the right).
}
   \label{fig:state-social-network-diagram}
\end{figure}

Building on the established link between SCI and real-world social connections, we introduce two variables, ``deaths in social proximity'' and ``deaths in spatial proximity'', to capture the influence of social and spatial factors on the distribution of OOD rates in counties in the United States. The term ``deaths in social proximity'' indicates the average number of death rates in alter locations weighted by their SCI to the focal node, also known as ``ego''. This variable operates as a socially lagged variable, accounting for the influence of death rates in ``alter'' locations, referring to the direct connections of the ego. On the other hand, ``deaths in spatial proximity'' measures the average number of death rates in the alters weighted by their inverse geographical distance to the ego. Unlike the socially lagged variable, this is a spatially lagged variable to account for the effects of deaths in the spatial vicinity of the ego. These factors work together to provide a comprehensive picture of how deaths are distributed and influenced by social and geographical factors in a given location. Quantitatively, deaths in social and spatial proximity, denoted by $s_{-i}$ and $d_{-i}$ for an ego location $i$, are defined as follows:
\begin{align*}
 s_{-i} = \sum_{j\neq i} w_{ij}y_{j},\; \mbox{ and } \; d_{-i} =\sum_{j\neq i} a_{ij}y_{j},
 \end{align*} where $y_i$ is OOD rate in US county $i$. The social and spatial proximity weights ($w_{ij}$ and $a_{ij}$) are defined as follows: \begin{align*}
 w_{ij} = \frac{n_{j}\mbox{SCI}_{ij}}{\sum_{k\neq i} n_{k} \mbox{SCI}_{ik}}, \; \mbox{ and } \;     
a_{ij} = \frac{1 + \frac{1}{d_{ij}}}{\sum_{k \neq i} (1 + \frac{1}{d_{ik}})}, 
\end{align*} where $n_j$ is the population of county $j$ and $d_{ij}$ is the distance between county locations $i$ and $j$. Using $1+1/d_{ij}$ instead of $1/d_{ij}$ in the definition of  $a_{ij}$ improves numerical stability when dealing with long distances (large $d_{ij}$ values) but does not change the decreasing nature of the spatial proximity weights with the increasing distances. To capture the effect of far away counties that would be otherwise down-weighted heavily by $1/d_{ij}$, in Supplementary Information section \ref{sec:distance decay weight check} we repeat our analysis with a slowly decaying distance function ($1/d^{^{1/10}}_{ij}$) and observe that our main conclusions remain unchanged and are not sensitive to the choice of the distance decay function. To visualize the social and spatial adjacency weights across the United States, we aggregate the county-level data to the state level. This aggregation allows us to effectively depict the dense networks. We give the details of the state level aggregation in Supplementary Information section \ref{sec:county maps}. It is important to note that our analysis of estimating the effect of ``deaths in social proximity'' on OODs is still conducted at the county level. The visualization of the proximity weights and proximity values, however, is presented at the state level to help understand the geographical dispersion of the social and spatial proximity weights in relation to OOD rates. For comparison, in Supplementary Information section \ref{sec:county maps}  we give the county-level visualization of social and spatial adjacency weights in Pennsylvania (PA). Figure \ref{fig:state-social-network-diagram}A illustrates the state-level social network, measured by the social proximity weights ($w_{xz}$) in the contiguous United States. Figures \ref{fig:state-social-network-diagram}B and \ref{fig:state-social-network-diagram}C show the spatial dispersion of the proximity weights for the socially and spatially lagged variables for two ego states, California (\ref{fig:state-social-network-diagram}B) and Pennsylvania (\ref{fig:state-social-network-diagram}C), in the contiguous US. 

\begin{figure}[!hbt]
   \centering
   \includegraphics[width=\textwidth]{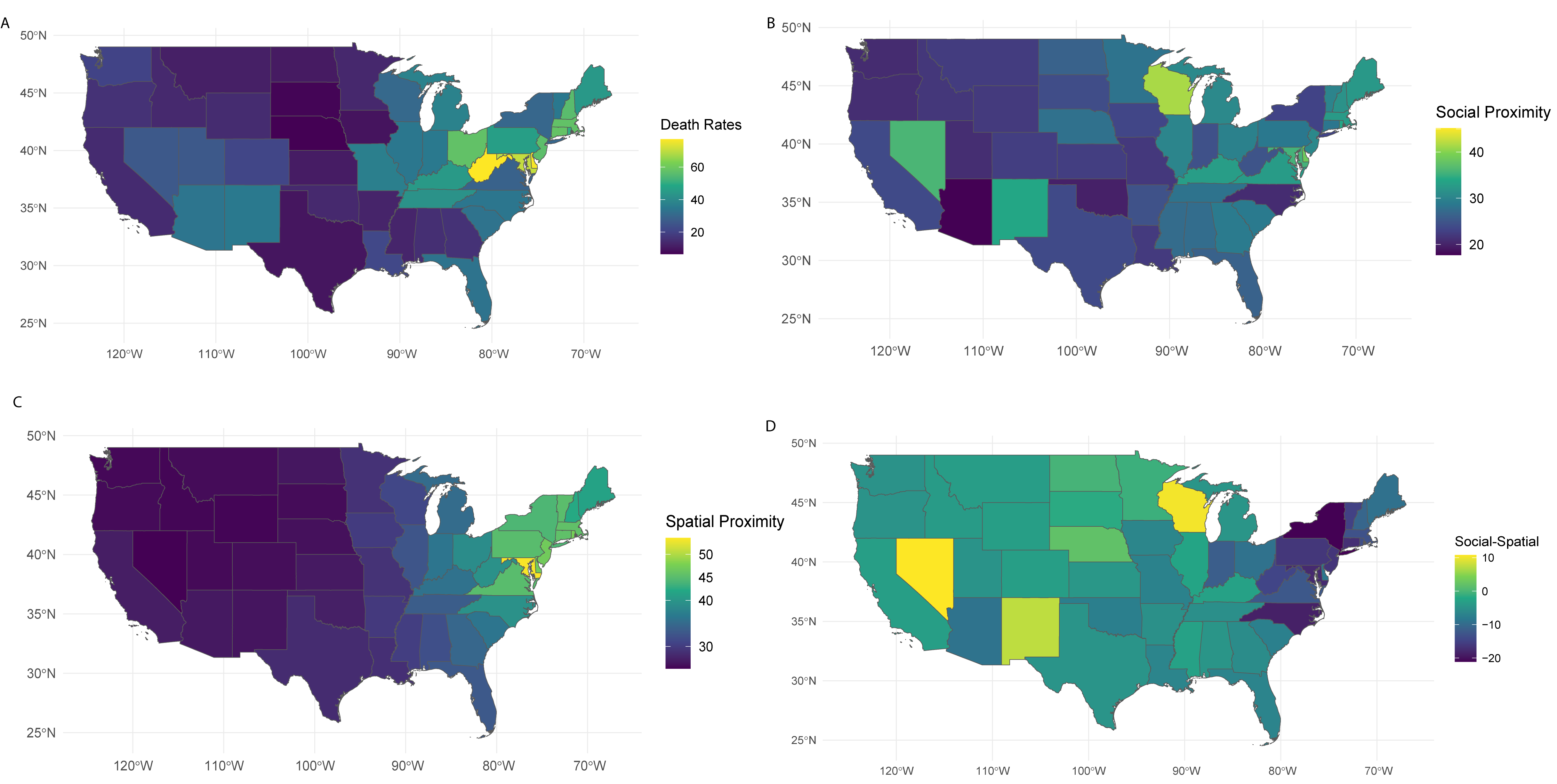}
   \caption{A) The top left map shows the spatial spread of state-level opioid overdose death rates in the contiguous US. B) The top right map shows the spatial dispersion of ``deaths in social proximity'' for states in the contiguous US. C) The bottom left map shows the geographical spread of ``deaths in spatial proximity''. D) The bottom right map shows the difference between deaths in social and spatial proximity from top right and bottom left maps.}
   \label{fig:State-Spatial-spread-of-primary-variable}
\end{figure}

\begin{figure}[!hbt]
   \centering
   \includegraphics[width=\textwidth]{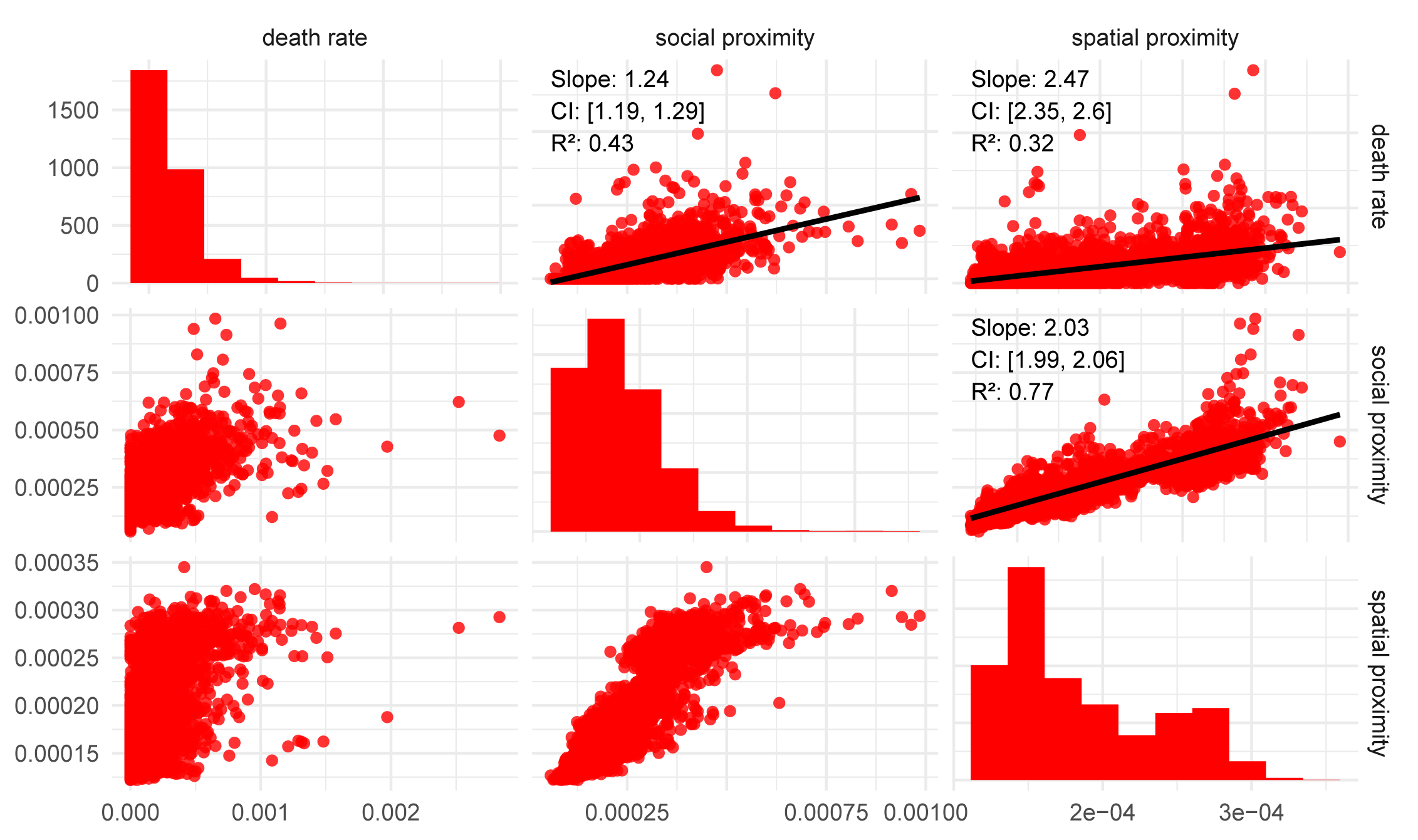}
   \caption{The figure shows the distribution and relationships between the primary variables of interest (death rates $y_{i}$, deaths in social proximity $s_{-i}$, and deaths in spatial proximity $d_{-i}$). The histograms on the main diagonal depict the distributions of $y_i$, $s_{-i}$, and $d_{-i}$. Moving to the upper triangle, we observe  the degree of linear dependence between these variables, while the lower triangle displays scatter plots.}
   \label{fig:scatter-plot-matrix}
\end{figure}

Having formally introduced the socially and spatially lagged variables, we use mortality data from the National Center for Health Statistics (NCHS) for 2018-2019
to measure the state-level OOD rates. Figure \ref{fig:State-Spatial-spread-of-primary-variable} shows the state-level spatial distributions of our main variables of interest in the contiguous US: OOD rates (Figure \ref{fig:State-Spatial-spread-of-primary-variable}A), deaths in social proximity (Figure \ref{fig:State-Spatial-spread-of-primary-variable}B), deaths in spatial proximity (Figure \ref{fig:State-Spatial-spread-of-primary-variable}C), and their differences (Figure \ref{fig:State-Spatial-spread-of-primary-variable}D). Figure \ref{fig:scatter-plot-matrix} shows the scatter plots of death rates ($y_i$), deaths in social proximity ($s_{-i}$), and deaths in spatial proximity ($d_{-i}$) for all counties in the contiguous US. The scatter plot matrix reveals a moderate linear dependence between death rates and spatially and socially lagged variables. In addition, there is a strong correlation between spatial and social proximity. However, the histograms that represent the distribution of these two variables exhibit differences. This contrast helps us identify the spatial effects of social influence on OOD and estimate an effect size for social proximity using controlled regressions described in the Results. Our choice of controls comes from domain knowledge consisting of clinical covariates and factors of social determinants of health (SDOH). The SDOH covariates are selected using the least absolute shrinkage and selection operator (LASSO) from an array of 17 variables. The details of SDOH and clinical covariates are in Methods.

\section*{Results}
\subsection*{Estimating the effect size of ``Deaths in Social Proximity'' variable}

\begin{figure}[!hbt]
   \centering
   \includegraphics[width=\textwidth]{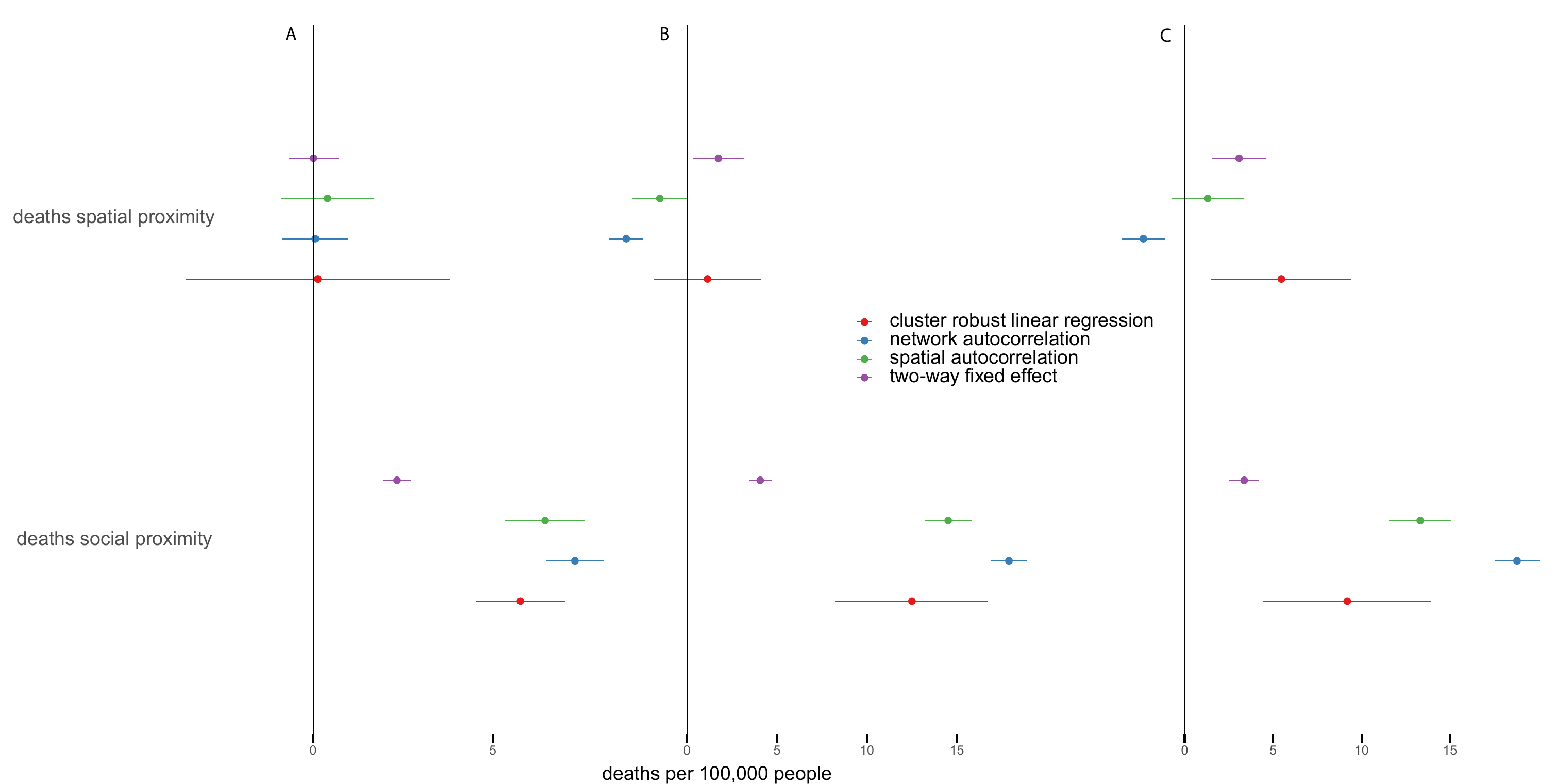}
   \caption{ A) The plot shows the coefficient confidence interval plots for western and central US counties. The coefficient for $s_{-i}$ for cluster-robust linear regression (Supplementary Table \ref{tab:simple_linear_model_with_robust_std_clustered_error-western-united-states}), network and spatial autocorrelation (Supplementary Table \ref{tab:network_spatial_autocorrelation_model_for_western_united_States}), and two-way fixed effects model (Supplementary Table \ref{tab:two-way-fixed-effect-western-united-states}), all indicate a positive, significant (p$<$0.001) coefficient for \(s_{-i}\). B). Shows the coefficient plot for social and spatial proximity for counties in the contiguous US. The coefficient for \(s_{-i}\) for cluster-robust linear regression (Supplementary Table \ref{tab:simple_linear_model_with_robust_std_clustered_error-contiguous-united-states}), network and spatial autocorrelation (Supplementary Table \ref{tab:network_spatial_autocorrelation_model_for_entire_united_States}), and two-way fixed effects models (Supplementary Table  \ref{tab:two-way-fixed-effect-entire-united-states}) are all positive and significant (p$<$ 0.001). C) Shows the coefficient plot for \(s_{-i}\) and \(d_{-i}\)  for counties in the eastern US. The coefficient for \(s_{-i}\) for cluster-robust linear regression (Supplementary Table \ref{tab:simple_linear_model_with_robust_std_clustered_error-eastern-united-state}), network and spatial autocorrelation models (Supplementary Table \ref{tab:network_spatial_autocorrelation_model_for_eastern_united_States}), and two-way fixed effects models (Supplementary Table \ref{tab:two-way-fixed-effect-eastern-united-states}) are all positive  and significant (p $<$ 0.001). The effect sizes for standardized $s_{-i}$ in the cluster-robust linear regression models indicate that a one-standard-deviation increase, equal to $11.69523$, $12.2417$, and $5.7145$ more deaths per $100,000$ population in the social proximity of the ego counties in contiguous, eastern and western-central United States, respectively, is associated with thirteen more deaths per $100,000$ population in contiguous and nine more deaths eastern and six more deaths in western-central US counties.}
   \label{fig:NBR-robustness-Coef-plot}
\end{figure}

Our outcome of interest is the county-level OOD rates that we measure using NCHS data from 2018 to 2019. We use cluster-robust linear regression to estimate the coefficient of the socially lagged variable. Robust estimators and clustering by state help us correct for correlation among counties, which might be higher for counties in the same state than between different states. Figure \ref{fig:NBR-robustness-Coef-plot} shows the significant coefficient of the socially lagged variable and provides statistical evidence for the effect of social networks on the spatial spread of OOD rates. The estimate of the effect size of the socially lagged variables is statistically significant across the eastern, western-central, and the entire contiguous US. The positive coefficient aligns with the theoretical proposition of the literature on the importance of social influence in the opioid epidemic \cite{costello2021peer}. The positive and significant magnitude of this effect size can originate from the dissemination of information through social networks about the initiation of use and availability of substances, leading to more OODs. The effect sizes for $s_{-i}$ and $d_{-i}$, derived from the cluster-robust standard error model, along with other covariates, are in Supplementary Tables \ref{tab:simple_linear_model_with_robust_std_clustered_error-eastern-united-state}, \ref{tab:simple_linear_model_with_robust_std_clustered_error-western-united-states}, and \ref{tab:simple_linear_model_with_robust_std_clustered_error-contiguous-united-states} for the eastern, western-central, and contiguous US. 

We standardize $s_{-i}$ and $d_{-i}$ prior to regression analysis. Consequently, the effect size for $s_{-i}$ indicates that an increase of one standard deviation, equal to $11.69523$ and $12.2417$ more ``deaths in social proximity'' per $100,000$ population in the contiguous and eastern United States, respectively, is associated with an increase of nine deaths per $100,000$ population in ego counties within the eastern United States and thirteen deaths per $100,000$ population in the contiguous United States. For counties in the western and central United States, a similar increase of one standard deviation, equal to $5.7145$ more deaths per $100,000$ population in the social proximity of ego counties, corresponds to six more deaths per $100,000$ population in ego counties. Despite observing a significant effect for the socially lagged variable on OOD rates, it is crucial to address potential issues arising from the inherent nature of the primary variable of interest. We recognize the methodological challenges due to correlated residuals when using statistical models to analyze social and spatial effects and perform robustness checks using network and spatial autocorrelation, as well as fixed-effects models to substantiate our findings. 

Given that lag variables have a social and spatial component, we expect the error terms in our regression model to be correlated. To address this, we implemented a spatial error model (SEM) to test and correct for network and spatial autocorrelations in error terms. The autocorrelation in error terms arises from unobserved factors shared among spatially or socially connected units. Cluster-robust linear regression may fail to capture these autocorrelation stemming from unobserved factors, which may lead to bias and inefficient estimates. Hence, to add robustness to our analysis, we utilize spatial and network aurocorrelation models. The methodological frameworks for the network and spatial autocorrelation models are explained in Supplementary Information sections \ref{sec:network_autocorrelation} and \ref{sec:spatial_autocorrelation}. Supplementary Tables \ref{tab:network_spatial_autocorrelation_model_for_eastern_united_States}, \ref{tab:network_spatial_autocorrelation_model_for_western_united_States} and \ref{tab:network_spatial_autocorrelation_model_for_entire_united_States} provide the SEM regression results for the network and spatial autocorrelation models for the eastern, western-central and contiguous United States, in which we find statistical evidence for correlated error. It is important to note that we performed two distinct models to test and correct for correlated error terms originating from both spatial and network dependence separately. Network autocorrelation might come from the structure of the socially lagged variable $s_{-i}$, while spatially correlated error terms could be attributed to the spatially lagged variable $d_{-i}$. Our autocorrelation models provide statistical evidence for the significance of social proximity in OODs.

To address spatial and temporal heterogeneity and enhance robustness, we employed a two-way fixed-effects model. Specifically, we included state-fixed effects to control for unobserved state-specific characteristics, such as regulations and policy environments, that are constant over time, but vary between states. We also incorporate year-fixed effects to absorb nationwide shocks or trends that could influence our outcome of interest. This modeling strategy enables us to robustly assess the statistical significance of ``deaths in social proximity'' while accounting for unobserved spatial and temporal confounders. The results of this model are in Supplementary Tables \ref{tab:two-way-fixed-effect-eastern-united-states}, \ref{tab:two-way-fixed-effect-western-united-states} and
\ref{tab:two-way-fixed-effect-entire-united-states} for the eastern, western-central and contiguous United States. Our findings consistently demonstrate a significant positive effect size for ``deaths in social proximity''. For a detailed explanation of the two-way fixed effect model, refer to Supplementary Information section \ref{sec:two-way-fixed-effect}.


Figure \ref{fig:NBR-robustness-Coef-plot} shows the confidence intervals (CI) for the cluster-robust standard error linear model, network autocorrelation, spatial autocorrelation, and the two-way, fixed-effects models. We consistently observe a positive and significant coefficient for $s_{-i}$, indicating the effect of social influence on the spread of OOD, while the effect size for $d_{-i}$ has a varying CI, changes sign, and is not always significant. We also observe cluster-robust linear regression has a broader confidence interval compared to the other models, it is primarily because cluster-robust standard errors adjust for the intra-cluster correlation by accounting for the fact that there is less independent information than the total number of observations suggests. This adjustment often results in larger standard errors compared to conventional ones, reflecting the reduced amount of independent information. There is also a loss of statistical power when doing this analysis, as the effective sample size becomes closer to the number of clusters rather than the total number of observations. However, in spite of accounting for intra-cluster correlation, we observe a statistically significant coefficient for ``deaths in social proximity''. A key takeaway from this result is that social connections are predominantly more significant than the effect of spatial proximity on OOD rates. This robustness check adds to the consistency of our statistical evidence for the size of the effect of ``deaths in social proximity''. In our analysis, given the network and spatial configurations inherent in $s_{-i}$ and $d_{-i}$, we suspected and addressed the correlated errors that stem from the endogeneity of both variables simultaneously, using a two-stage least squares approach. The results of our implementation are discussed in the Supplementary Information section \ref{sec:G2sls} and confirm the positive and significant effect size of our social proximity variable in the eastern, western-central and entire contiguous United States (Supplementary Figure \ref{fig:G2SLS-Coef-plot} and Supplementary Table \ref{tab:G2SLS-east-west-entire_us}). To end our analysis, we show the robustness of the effect of ``deaths in social proximity''  by accounting for the rate at which spatial adjacency weights decay with increasing distance. By accounting for the ``deaths in spatial proximity'' decay rate, we allow the effect of distant counties far from the focal county to be more pronounced. We use cluster-robust linear regression to test the significance of the coefficient $s_{-i}$. We observe a statistically significant effect for the coefficient of ``deaths in social proximity'' between counties in the eastern, western-central, and contiguous United States. The results of this implementation are discussed in the Supplementary Information Section \ref{sec:distance decay weight check}. The confidence interval plot for the model is illustrated in Supplementary Figure \ref{fig:lm-sw-Coef-plot} and the effect sizes are shown in Supplementary Tables \ref{tab:linear_regression_eastern_united_states_distance_decay_weight}, \ref{tab:linear_regression_western_united_states_distance_decay_weight}, and \ref{tab:linear_regression_entire_united_states_distance_decay_weight}. 


\section*{Discussion}


 Our research underscores that social influence, measured by the coefficient of $s_{-i}$, exhibits a statistically significant impact on the spatial spread of OOD rates in US counties. This finding paves the way for a more in-depth exploration of the mechanisms driving the opioid epidemic, specifically within the social contagion framework. Harmon et al. (2015) \cite{harmon2015anticipating} and Braha and Aguiar (2017) \cite{braha2017voting} have utilized complex generative modeling to understand social contagion in the context of economic crises and voting behavior. Building on their work in modeling social contagion, future studies can adopt a similar framework to model peer influence in the opioid epidemic using SCI data. 
 
 Our analysis demonstrates that deaths in social proximity, as measured by the SCI, are associated with opioid overdose deaths in counties in the United States. Although previous studies recognize the tendency for OODs to occur in isolated spaces \cite{gicquelais2022prevalence,li2022understanding}, data from the CDC's State Unintentional Drug Overdose Reporting System (SUDORS)\footnote{\url{https://www.cdc.gov/overdose-prevention/data-research/facts-stats/sudors-dashboard-fatal-overdose-data.html}} indicate that almost half of overdose deaths had a potential bystander present, suggesting that the occurrence of OODs in isolation may be less pronounced than previously thought. 
 
 Furthermore, our measure of social influence is based on the social connections derived from the social network structure of Facebook's SCI across US counties and serves as a proxy for aggregated friendship network structures and their association with OODs; however, it does not capture the strength of individual friendship ties among Facebook users in the counties. Focusing on these social connections and their relation to the spatial patterns of OODs, such as geographic discordance (the community in which overdose death occurs being different from the community of residence), can provide novel insights into the complex interplay of social and spatial factors in perpetuating this public health crisis. 

In addressing the opioid epidemic, accounting for social networks is crucial in understanding and predicting OOD patterns at the community level to aid policy intervention.  Studies using complex generative models with spatial clustering have yielded valuable insight into the dynamics of social influence and can be crucial to designing targeted interventions. For example, Braha and Aguiar (2017) \cite{braha2017voting} use such models to identify spatial clusters of voting contagion. In the context of the opioid epidemic, Liao et al. (2024) \cite{liao2022tides} introduce the Spatio-TEMporal Mutually Exciting Point Process (STEMMED) model to quantify the interconnections between historical and future events, reflecting space-time clustering in OODs across various communities. This methodological approach can improve the prediction of OOD trends within localized settings by modeling unique community-specific OOD event streams, considering spatial and social influences. Our research complements this perspective by bridging the gap in understanding the role of social influence within these dynamics. Although STEMMED captures the spatiotemporal clustering of OOD events, our approach highlights the importance of social structures in shaping these patterns. By estimating the coefficient of the socially lagged variable, we measure the effect of social interactions in different counties in the US, which in conjunction with the spatial focus of Liao et al. \cite{liao2022tides} can offer a more comprehensive view of the factors driving OOD clusters. For policymakers, this comprehensive approach can provide a solid foundation for designing targeted interventions that address both the spatial and social dimensions of the opioid epidemic.


One of our motivations for this study was to facilitate the creation of proxy networks for social interactions in agent-based models of the US population based on SCI data. Agent-based models with SCI-calibrated social networks can provide valuable information about peer effects in the initiation of opioid use and the development of use disorder. When used in conjunction with powerful agent-based modeling tools such as FRED (Framework for Reconstructing Epidemic Dynamics) \cite{grefenstette2013fred}, such models can enable epidemiologists and policymakers to simulate the spread of the opioid epidemic in great mechanistic detail and evaluate interventions. Chu et al. (2020) use FRED to study social contagion in the use of e-cigarettes \cite{chu2020integrating}, showing the potential of FRED to simulate social contagion of addictive behaviors. We expect that SCI-calibrated social networks, combined with agent-based models of OUD progression in FRED (cf., e.g., \cite{ahmed2023estimating}), can reveal important mechanistic details about the opioid epidemic in the US. Our results may be particularly useful for calibrating social contagion parameters. The contagion parameters can be tuned so that they reflect the same association between ``deaths in social proximity'' and OODs as we measure in the mortality data.

SCI can also help design and evaluate intervention strategies based on social networks. Macmadu et al. (2022) highlight the need for interventions tailored to social networks \cite{macmadu2022drug}. Their research shows that after an overdose, network members exhibit observable behavioral changes, including increased risk taking to manage feelings of bereavement and trauma, protective actions, and some cases showing no change in drug use behavior. Based on these findings, opioid reduction efforts can be optimized in areas with greater social connectivity. Measurement of the effects of social networks across locations can also help optimize the allocation of limited medications, such as naloxone, buprenorphine, and methadone. In all these cases, social network-based strategies offer important opportunities for policy makers to mitigate the spread of the opioid epidemic.

 We acknowledge the limitations of using mortality data to measure the spread of this critical health crisis and propose to strengthen these findings with other OUD-related outcomes. Details on the limitations of the mortality data are discussed in the Methods. Future work can use SCI to investigate the social contagion of the opioid epidemic in mechanistic agent-based models and structural causal models for causal discovery and inference. We also plan to use SCI to improve our predictive accuracy in detecting OOD hotspots in the US. We have used SCI to measure the dynamics of social networks and their effect on the opioid epidemic. Our choice of SCI is guided by existing literature and our objective to study network effects at the population scale. We recognize the limitations of using SCI because Facebook users are not the same as the general population. SCI only acts as a proxy for real-life friendship networks, and it may not be a true representative of real-life friendship connections and their association with opioid use behavior. Other datasets that capture real-world interactions within a social network framework can also be valuable to understand the effects of social networks in the opioid epidemic. For example, the Add Health dataset \footnote{\url{https://data.cpc.unc.edu/projects/2/view\#public_li}}, which includes information on best friends during high school and opioid misuse in adulthood (14 years later), offers an alternative means of measuring these effects. Although this dataset allows analysis within the age-specific bracket of 25 to 35 years, potential problems of population selection bias may limit the generalizability of the findings to the broader population.

\section*{Methods}
\subsection*{Data pipeline and preprocessing} 

\begin{figure}[thb]
   \centering
   \includegraphics[width=\textwidth]{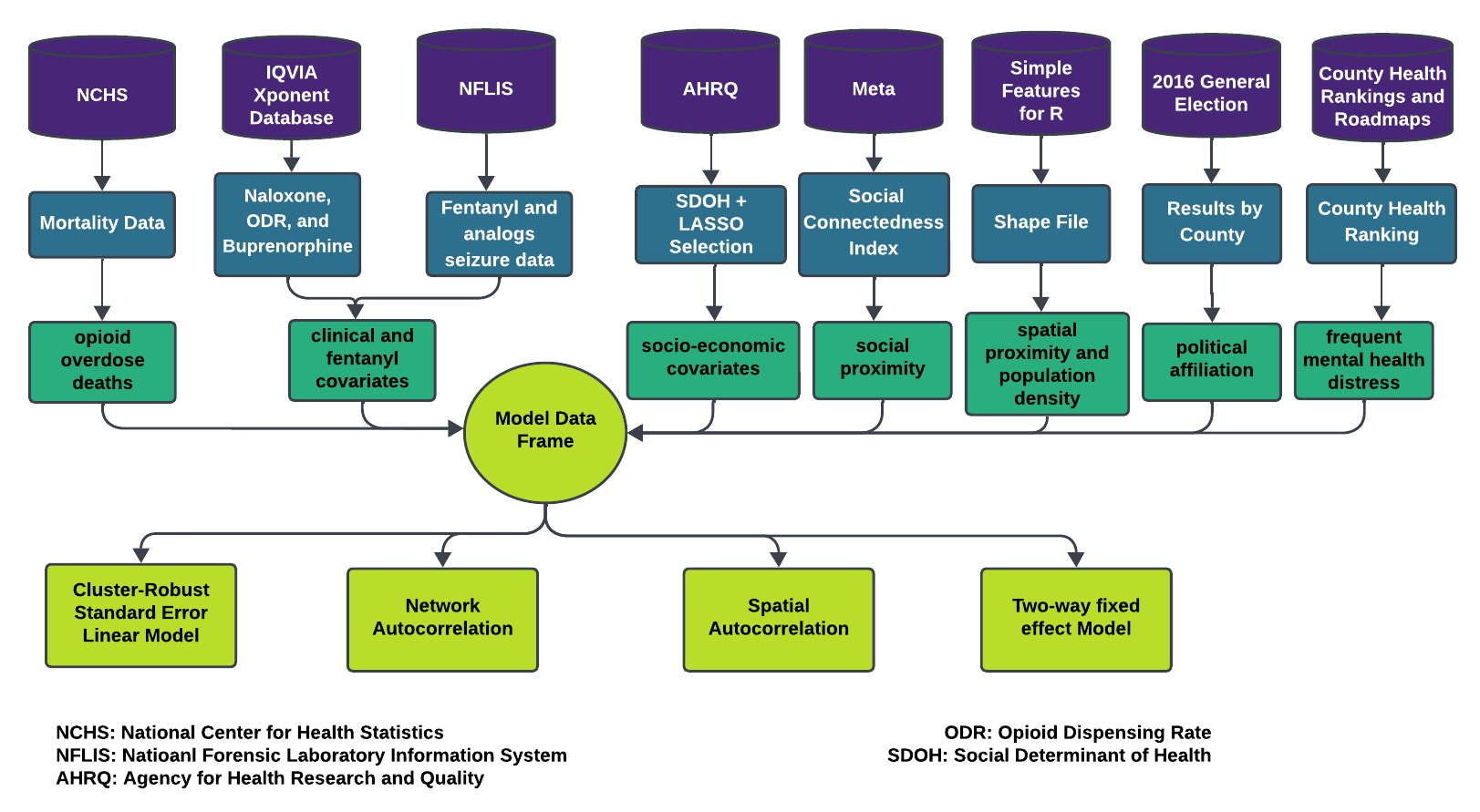}
   \caption{The diagram depicts the data pipeline for our analysis, including the data streams for the primary variable of interests, $s_{-i}$ and $d_{-i}$, as well as the relevant socioeconomic and clinical covariates. It also outlines our regression models for estimating the effect of peer influence as measured by SCI on county-level OOD outcomes.}
   \label{fig:datapipeline-plot}
\end{figure}

Our data sources are shown in Figure \ref{fig:datapipeline-plot} and explained in detail in the following paragraphs.

\paragraph{Mortality data and census demographics.} We measure the OOD rate from mortality data obtained from the National Center for Health Statistics (NCHS) for the years 2018-2019. This data set includes demographic details of individuals who have lost their lives to opioid-related overdoses. To first identify overdose related deaths we utilise the following International Classification of Disease (ICD) codes ``X40",``X41", ``X42", ``X43", ``X44",``X60", ``X61", ``X62",``X63", ``X64", ``X85", ``Y10", ``Y11", ``Y12", ``Y13", ``Y14". The X and Y codes provide information about deaths that have occurred due to substance overdose. Furthermore, to specifically target opioid overdose deaths we use the ICD T codes ``T400", ``T401", ``T402", ``T403", ``T404", and ``T406". The T codes determine the cause of death in the specification of opioids from other substances. Despite the comprehensive nature of the NCHS mortality data and the use of specific ICD codes to identify opioid overdose deaths, there are important limitations associated with this dataset. Death certificates may not always specify the drugs involved in an overdose, and some overdose deaths involve multiple drugs, making it difficult to determine which substance was primarily responsible. In addition, the analysis also incorporates demographic data on the broader population of the United States of America. The data is stratified at the county levels. To extract this information, we utilize the R package ``tidycensus" to systematically retrieve data from data.census.gov.

\paragraph{Clinical and mental health covariates.} Clinical factors such as the opioid dispensing rate (ODR), availability of naloxone, and access to buprenorphine for the treatment of opioid use disorder are used as controls in our regression. Morgan et al. (2018) \cite{morgan2018role} underscore the role of naloxone in reducing opioid-related harm, while Pendergrass et al. (2019) \cite{pendergrass2019importance} emphasize the importance of buprenorphine availability in mitigating overdose fatalities. The data source for clinical factors such as naloxone, buprenorphine, and ODR is the IQVIA Xponent database, which provides the number of prescriptions for naloxone and buprenorphine distributed throughout counties in the US through retail pharmacy channels. The ODR, which captures the total number of opioid doses dispensed, is measured as morphine milligram equivalents (MME). The current wave of the opioid epidemic from 2013 to the present has witnessed an increase in the illicit use of synthetic opioids such as fentanyl. Furthermore, Kuehn (2023) \cite{kuehn2023fentanyl} discuses the impact of fentanyl on overdose deaths, particularly among adolescents. Pergolizzi et al. (2018) \cite{pergolizzi2018going} also discuss the role of fentanyl in exacerbating the opioid epidemic. To control for the supply of illegal substances that contribute to OODs, we incorporate state-level fentanyl and analog seizure data from the National Forensic Laboratory Information System as a control variable. Our selection of clinical and illicit-supply covariates is comprehensive based on the data sources available to us. In addition, several studies have shown associations between mental health and opioid overdose deaths \cite{dinwiddie2024reported, santo2022prevalence, van2022mental}. Thus, we used data on frequent mental health distress from County Health Rankings and Roadmaps (CHRR). Frequent mental distress is the percentage of adults who reported poor mental health for more than 14 days in response to the question ``How many days during the past 30 days was your mental health not good?'' \cite{Yang2019}. We use this measure to control for the effect of mental health-related issues in counties. 
 
 \paragraph{Population density and political affiliation.} We also control for population density and political affiliation that are identified in the literature on the opioid epidemic and the structure of social networks. To account for the heterogeneity associated with SCI and opioid use among populations residing in urban and rural counties, we include population density in our regression model \cite{bailey2020social, peters2020opioid}. The risk status of opioid misuse is also associated with the political affiliations and liberal status of states \cite{haffajee2019characteristics}. Therefore, we control for the effect of political affiliation in our regression by accounting for counties' political leanings using 2016 general election data at the county level. 

\paragraph{SCI data.} SCI is available at the ZIP code and county levels in the United States. We chose counties as our analysis unit instead of ZIP codes because the latter had significant limitations. A considerable number of ZIP codes have missing SCI data. Typically, ZIP codes without SCI data are those with low populations or those designated exclusively for institutions. Institutional ZIP codes are assigned to areas predominantly occupied by specific institutions, such as hospitals, universities, or military bases. These institutional ZIP codes can also introduce spatial bias; for example, ZIP codes with hospitals are more likely to report higher overdose death rates. To ensure a more continuous and representative spatial framework, we use counties as our unit of analysis. This approach mitigates the issues of missing data and spatial bias, providing a more robust basis for analysis.

\paragraph{Social determinants of health.} Socioeconomic factors shape social structures and ties; therefore, incorporating these factors is essential to interpret the influence of social networks on OOD. Shared socioeconomic conditions can also influence behaviors that mirror peer influence, and therefore, we include socioeconomic covariates as controls in our analysis to mitigate the risk of bias due to missing confounders when estimating the effect size of our deaths in the social proximity variable ($s_{-i}$). Social determinants of health (SDOH) encompass aspects of physical infrastructure, economic context, healthcare context, and social environment of counties. Our selection of SDOH covariates is based on socioeconomic predictors of the opioid epidemic. Liu et al. (2023) \cite{liu2023geo} demonstrate the effects of SDOH measures on drug overdose death locations. Therefore, we use SDOH variables to control for socioeconomic factors that may confound the impact of $s_{-i}$ on county-level OOD rates. These covariates are selected from a pool of SDOH variables using the Least Absolute Shrinkage and Selection Operator (LASSO) technique to avoid multicollinearity issues. We select a subset of covariates from an array of 17 SDOH variables that are listed in Supplementary Table \ref{tab:Merged_Variables_Coefficients}. We provide details of the LASSO selection process in the Supplementary Information section \ref{sec:variable-selection}.



\paragraph{Statistical models.} The model coefficients are evaluated for statistical significance at $p  < 0.05$ level. Using county-level opioid overdose death rates as the outcome variable, we set up regression models to test the significance of deaths in social proximity to explain the county-level death rate after controlling for deaths in spatial proximity, clinical (mental health, availability of naloxone and buprenorphine and opioid dispensing rates in pharmacies) and fentanyl covariates, and socioeconomic covariates selected using LASSO from SDOH variables, as well as population density and political affiliation covariates. Linear regression and cluster-robust linear regression are analyzed for statistical significance (details in Supplementary Information section \ref{sec:section_linear_regression}). The residuals in the regression are weighted by population size of the counties, so that the inference is reflecting the population residing in the counties across the US. To account for correlated error terms from the spatial and network structure of the variables $d_{-i}$ and $s_{-i}$, respectively, we use spatial and network autocorrelation models (details in Supplementary Information sections \ref{sec:spatial_autocorrelation} and \ref{sec:network_autocorrelation}). To account for unobserved space- and time-invariant characteristics that might be associated with the covariates in our regression, we use a two-way fixed-effects model. This model includes fixed temporal effects for 2018 and 2019 and fixed spatial effects for states in the eastern United States, the western and central United States, and the contiguous United States (details in Supplementary Information section \ref{sec:two-way-fixed-effect}). As a robustness check, we also use a two-stage least squares regression to account for correlated error terms that arise from the simultaneous estimation of effects from $s_{-i}$ and $d_{-i}$ (details in the Supplementary Information section \ref{sec:G2sls}). As a final robustness check, we use distance decay weight to account for the effect of faraway counties from the focal county (details in the Supplementary Information section \ref{sec:distance decay weight check}). 

\section*{Acknowledgements}
 The project is supported by the contract 75D30121C12574 from the Centers for Disease Control and Prevention (CDC). The findings and conclusions in this work are those of the authors and do not represent the official position of the CDC. Along with this, the University of Pittsburgh Center for Research Computing, RRID: SCR 022735, has supported our work. This work utilizes the H2P cluster, supported by NSF award number OAC-2117681. The funders had no role in study design, data collection and analysis, decision to publish or preparation of the manuscript. The authors also acknowledge the contributions of Alexander Hayes, Nisha Natraj, William Wei Wang, and Kun Zhang for their valuable discussions and constructive feedback. 
\section*{Data Availability}
Mortality data was obtained from the National Center for Health Statistics (NCHS). Due to confidentiality concerns, this data set is not publicly accessible, but can be requested from NCHS at \url{https://www.cdc.gov/nchs/nvss/nvss-restricted-data.htm}. The clinical covariates were sourced from the IQVIA Xponent database, which is also not publicly available. Access can be requested through IQVIA at \url{https://www.iqvia.com/insights/the-iqvia-institute/available-iqvia-data}. Data on illicit fentanyl-related drugs were obtained from the National Forensic Laboratory Information System (NFLIS) and can be accessed at \url{https://www.nflis.deadiversion.usdoj.gov/}. Data on frequent mental health distress are obtained from the County Health Rankings and Roadmaps (CHRR) at \url{https://www.countyhealthrankings.org/sites/default/files/media/document/analytic_data2019}. The 2016 General Election data can be accessed at \url{https://raw.githubusercontent.com/tonmcg/US_County_Level_Election_Results_08-20/master/2016_US_County_Level_Presidential_Results}. The social determinants of health covariates are available from the Agency for Healthcare Research and Quality (AHRQ) at \url{https://www.ahrq.gov/sdoh/data-analytics/sdoh-data.html}. The SCI data can be accessed using the Facebook Data for Good tools at \url{https://dataforgood.facebook.com/dfg/tools/social-connectedness-index}. Data on drug overdose deaths used in our LASSO covariate selection are available from the Bureau of Health Statistics and Registries of the Department of Health of the Commonwealth of Pennsylvania and can be requested as follows:
\url{https://www.health.pa.gov/topics/Documents/Reporting-Registries/VR-Govt-Researchers/Application.%20for%20Access%20to%20Protected%20Data%20for%20Public%20Health%20Researchers%20-%20User%27s%20Guide.pdf}.

\section*{Code Availability}
Analysis code to reproduce figures and tables in the paper is available at \url{https://github.com/kut97/ood-sci}
\bibliography{refs.bib}

\begin{thebibliography}{10}

\bibitem{jalal2018changing}
H.~Jalal, J.~M. Buchanich, M.~S. Roberts, L.~C. Balmert, K.~Zhang, and D.~S. Burke, ``Changing dynamics of the drug overdose epidemic in the {United States} from 1979 through 2016,'' {\em Science}, vol.~361, no.~6408, p.~eaau1184, 2018.

\bibitem{jalal2022exponential}
H.~Jalal and D.~S. Burke, ``Exponential growth of drug overdose poisoning and opportunities for intervention,'' {\em Addiction}, vol.~117, no.~5, pp.~1200--1202, 2022.

\bibitem{alpert2022origins}
A.~Alpert, W.~N. Evans, E.~M. Lieber, and D.~Powell, ``Origins of the opioid crisis and its enduring impacts,'' {\em The Quarterly Journal of Economics}, vol.~137, no.~2, pp.~1139--1179, 2022.

\bibitem{lim2022modeling}
T.~Y. Lim, E.~J. Stringfellow, C.~A. Stafford, C.~DiGennaro, J.~B. Homer, W.~Wakeland, S.~L. Eggers, R.~Kazemi, L.~Glos, E.~G. Ewing, {\em et~al.}, ``Modeling the evolution of the {US} opioid crisis for national policy development,'' {\em Proceedings of the National Academy of Sciences}, vol.~119, no.~23, p.~e2115714119, 2022.

\bibitem{costello2021peer}
B.~J. Costello, B.~J. Anderson, and M.~Stein, ``Peer influence in initiation to heroin use,'' {\em Journal of Drug Issues}, vol.~51, no.~2, pp.~323--339, 2021.

\bibitem{rigg2018opioid}
K.~K. Rigg, K.~McLean, S.~M. Monnat, G.~E. Sterner~III, and A.~M. Verdery, ``Opioid misuse initiation: implications for intervention,'' {\em Journal of addictive diseases}, vol.~37, no.~3-4, pp.~111--122, 2018.

\bibitem{guarino2018young}
H.~Guarino, P.~Mateu-Gelabert, J.~Teubl, and E.~Goodbody, ``Young adults' opioid use trajectories: From nonmedical prescription opioid use to heroin, drug injection, drug treatment and overdose,'' {\em Addictive behaviors}, vol.~86, pp.~118--123, 2018.

\bibitem{jalal2020age}
H.~Jalal, J.~M. Buchanich, D.~R. Sinclair, M.~S. Roberts, and D.~S. Burke, ``Age and generational patterns of overdose death risk from opioids and other drugs,'' {\em Nature medicine}, vol.~26, no.~5, pp.~699--704, 2020.

\bibitem{syvertsen2017down}
J.~L. Syvertsen, C.~E. Paquette, and R.~A. Pollini, ``Down in the valley: Trajectories of injection initiation among young injectors in california’s central valley,'' {\em International Journal of Drug Policy}, vol.~44, pp.~41--49, 2017.

\bibitem{valente2007peer}
T.~W. Valente, A.~Ritt-Olson, A.~Stacy, J.~B. Unger, J.~Okamoto, and S.~Sussman, ``Peer acceleration: effects of a social network tailored substance abuse prevention program among high-risk adolescents,'' {\em Addiction}, vol.~102, no.~11, pp.~1804--1815, 2007.

\bibitem{stringfellow2022reducing}
E.~J. Stringfellow, T.~Y. Lim, K.~Humphreys, C.~DiGennaro, C.~Stafford, E.~Beaulieu, J.~Homer, W.~Wakeland, B.~Bearnot, R.~K. McHugh, {\em et~al.}, ``Reducing opioid use disorder and overdose deaths in the {United States}: A dynamic modeling analysis,'' {\em Science Advances}, vol.~8, no.~25, p.~eabm8147, 2022.

\bibitem{braha2017voting}
D.~Braha and M.~A. De~Aguiar, ``Voting contagion: Modeling and analysis of a century of {US} presidential elections,'' {\em {PLOS} ONE}, vol.~12, no.~5, p.~e0177970, 2017.

\bibitem{homer2021dynamic}
J.~Homer and W.~Wakeland, ``A dynamic model of the opioid drug epidemic with implications for policy,'' {\em The American Journal of Drug and Alcohol Abuse}, vol.~47, no.~1, pp.~5--15, 2021.

\bibitem{bailey2018social}
M.~Bailey, R.~Cao, T.~Kuchler, J.~Stroebel, and A.~Wong, ``Social connectedness: Measurement, determinants, and effects,'' {\em Journal of Economic Perspectives}, vol.~32, no.~3, pp.~259--80, 2018.

\bibitem{bailey2020social}
M.~Bailey, P.~Farrell, T.~Kuchler, and J.~Stroebel, ``Social connectedness in urban areas,'' {\em Journal of Urban Economics}, vol.~118, p.~103264, 2020.

\bibitem{bailey2020determinants}
M.~Bailey, D.~Johnston, T.~Kuchler, D.~Russel, B.~State, and J.~Stroebel, ``The determinants of social connectedness in europe,'' in {\em Social Informatics: 12th International Conference, SocInfo 2020, Pisa, Italy, October 6--9, 2020, Proceedings 12}, pp.~1--14, Springer, 2020.

\bibitem{coven2023jue}
J.~Coven, A.~Gupta, and I.~Yao, ``{JUE Insight}: Urban flight seeded the {COVID-19} pandemic across the {United States},'' {\em Journal of Urban Economics}, vol.~133, p.~103489, 2023.

\bibitem{kuchler2022jue}
T.~Kuchler, D.~Russel, and J.~Stroebel, ``{JUE} insight: The geographic spread of {COVID-19} correlates with the structure of social networks as measured by facebook,'' {\em Journal of Urban Economics}, vol.~127, p.~103314, 2022.

\bibitem{ge2024reddit}
Y.~Ge, S.~Das, K.~O'Connor, M.~A. Al-Garadi, G.~Gonzalez-Hernandez, and A.~Sarker, ``Reddit-impacts: A named entity recognition dataset for analyzing clinical and social effects of substance use derived from social media,'' {\em arXiv preprint arXiv:2405.06145}, 2024.

\bibitem{pandrekar2018social}
S.~Pandrekar, X.~Chen, G.~Gopalkrishna, A.~Srivastava, M.~Saltz, J.~Saltz, and F.~Wang, ``Social media based analysis of opioid epidemic using reddit,'' in {\em AMIA Annual Symposium Proceedings}, vol.~2018, p.~867, American Medical Informatics Association, 2018.

\bibitem{balsamo2021patterns}
D.~Balsamo, P.~Bajardi, A.~Salomone, and R.~Schifanella, ``Patterns of routes of administration and drug tampering for nonmedical opioid consumption: data mining and content analysis of reddit discussions,'' {\em Journal of Medical Internet Research}, vol.~23, no.~1, p.~e21212, 2021.

\bibitem{diemer2022no}
A.~Diemer and T.~Regan, ``No inventor is an island: social connectedness and the geography of knowledge flows in the us,'' {\em Research Policy}, vol.~51, no.~2, p.~104416, 2022.

\bibitem{mackle2022social}
K.~M{\"a}ckle and S.~Ruenzi, ``Friends with drugs: The role of social networks in the opioid epidemic,'' {\em Available at SSRN}, 2022.

\bibitem{cutler2024thick}
D.~M. Cutler and J.~T. Donahoe, ``Thick market externalities and the persistence of the opioid epidemic,'' tech. rep., National Bureau of Economic Research, 2024.

\bibitem{forati2023journey}
A.~Forati, R.~Ghose, F.~Mohebbi, and J.~R. Mantsch, ``The journey to overdose: Using spatial social network analysis as a novel framework to study geographic discordance in overdose deaths,'' {\em Drug and alcohol dependence}, vol.~245, p.~109827, 2023.

\bibitem{koram2011role}
N.~Koram, H.~Liu, J.~Li, J.~Li, J.~Luo, and J.~Nield, ``Role of social network dimensions in the transition to injection drug use: actions speak louder than words,'' {\em AIDS and Behavior}, vol.~15, pp.~1579--1588, 2011.

\bibitem{xu2022beyond}
Y.~Xu, P.~Santi, and C.~Ratti, ``Beyond distance decay: Discover homophily in spatially embedded social networks,'' {\em Annals of the American Association of Geographers}, vol.~112, no.~2, pp.~505--521, 2022.

\bibitem{harmon2015anticipating}
D.~Harmon, M.~Lagi, M.~A. De~Aguiar, D.~D. Chinellato, D.~Braha, I.~R. Epstein, and Y.~Bar-Yam, ``Anticipating economic market crises using measures of collective panic,'' {\em PloS one}, vol.~10, no.~7, p.~e0131871, 2015.

\bibitem{gicquelais2022prevalence}
R.~E. Gicquelais, B.~L. Genberg, J.~L. Maksut, A.~S. Bohnert, and A.~C. Fernandez, ``Prevalence and correlates of using opioids alone among individuals in a residential treatment program in michigan: implications for overdose mortality prevention,'' {\em Harm reduction journal}, vol.~19, no.~1, p.~135, 2022.

\bibitem{li2022understanding}
Y.~Li, H.~J. Miller, E.~D. Root, A.~Hyder, and D.~Liu, ``Understanding the role of urban social and physical environment in opioid overdose events using found geospatial data,'' {\em Health \& Place}, vol.~75, p.~102792, 2022.

\bibitem{liao2022tides}
C.-Y. Liao, G.-G. Garcia, K.~Paynabar, Z.~Dong, Y.~Xie, and M.~S. Jalali, ``Tides need stemmed: A locally operating spatio-temporal mutually exciting point process with dynamic network for improving opioid overdose death prediction,'' {\em arXiv preprint arXiv:2211.07570}, 2022.

\bibitem{grefenstette2013fred}
J.~J. Grefenstette, S.~T. Brown, R.~Rosenfeld, J.~DePasse, N.~T. Stone, P.~C. Cooley, W.~D. Wheaton, A.~Fyshe, D.~D. Galloway, A.~Sriram, {\em et~al.}, ``Fred (a framework for reconstructing epidemic dynamics): an open-source software system for modeling infectious diseases and control strategies using census-based populations,'' {\em BMC Public Health}, vol.~13, pp.~1--14, 2013.

\bibitem{chu2020integrating}
K.-H. Chu, A.~Shensa, J.~B. Colditz, J.~E. Sidani, B.~L. Hoffman, D.~Sinclair, M.~G. Krauland, and B.~A. Primack, ``Integrating social dynamics into modeling cigarette and e-cigarette use,'' {\em Health Education \& Behavior}, vol.~47, no.~2, pp.~191--201, 2020.

\bibitem{ahmed2023estimating}
A.~A. Ahmed, M.~A. Rahimian, and M.~S. Roberts, ``Estimating treatment effects using costly simulation samples from a population-scale model of opioid use disorder,'' in {\em 2023 IEEE EMBS International Conference on Biomedical and Health Informatics (BHI)}, pp.~1--4, IEEE, 2023.

\bibitem{macmadu2022drug}
A.~Macmadu, L.~Frueh, A.~B. Collins, R.~Newman, N.~P. Barnett, J.~D. Rich, M.~A. Clark, and B.~D. Marshall, ``Drug use behaviors, trauma, and emotional affect following the overdose of a social network member: a qualitative investigation,'' {\em International Journal of Drug Policy}, vol.~107, p.~103792, 2022.

\bibitem{morgan2018role}
J.~Morgan and A.~L. Jones, ``The role of naloxone in the opioid crisis,'' {\em Toxicology Communications}, vol.~2, no.~1, pp.~15--18, 2018.

\bibitem{pendergrass2019importance}
S.~A. Pendergrass, R.~C. Crist, L.~K. Jones, J.~R. Hoch, and W.~H. Berrettini, ``The importance of buprenorphine research in the opioid crisis,'' {\em Molecular Psychiatry}, vol.~24, no.~5, pp.~626--632, 2019.

\bibitem{kuehn2023fentanyl}
B.~M. Kuehn, ``Fentanyl drives startling increases in adolescent overdose deaths,'' {\em JAMA}, vol.~329, no.~4, pp.~280--281, 2023.

\bibitem{pergolizzi2018going}
J.~V. Pergolizzi~Jr, J.~A. LeQuang, R.~Taylor~Jr, R.~B. Raffa, and N.~R. Group, ``Going beyond prescription pain relievers to understand the opioid epidemic: the role of illicit fentanyl, new psychoactive substances, and street heroin,'' {\em Postgraduate Medicine}, vol.~130, no.~1, pp.~1--8, 2018.

\bibitem{dinwiddie2024reported}
A.~T. Dinwiddie, ``Reported non-substance-related mental health disorders among persons who died of drug overdose—{United States}, 2022,'' {\em MMWR. Morbidity and Mortality Weekly Report}, vol.~73, 2024.

\bibitem{santo2022prevalence}
T.~Santo~Jr, G.~Campbell, N.~Gisev, D.~Martino-Burke, J.~Wilson, S.~Colledge-Frisby, B.~Clark, L.~T. Tran, and L.~Degenhardt, ``Prevalence of mental disorders among people with opioid use disorder: A systematic review and meta-analysis,'' {\em Drug and Alcohol Dependence}, vol.~238, p.~109551, 2022.

\bibitem{van2022mental}
J.~van Draanen, C.~Tsang, S.~Mitra, V.~Phuong, A.~Murakami, M.~Karamouzian, and L.~Richardson, ``Mental disorder and opioid overdose: a systematic review,'' {\em Social Psychiatry and Psychiatric Epidemiology}, pp.~1--25, 2022.

\bibitem{Yang2019}
T.-C. Yang, S.~A. Matthews, F.~Sun, and M.~Armendariz, ``Modeling the importance of within- and between-county effects in an ecological study of the association between social capital and mental distress,'' {\em Preventing Chronic Disease}, vol.~16, p.~E75, June 2019.

\bibitem{peters2020opioid}
D.~J. Peters, S.~M. Monnat, A.~L. Hochstetler, and M.~T. Berg, ``The opioid hydra: Understanding overdose mortality epidemics and syndemics across the rural-urban continuum,'' {\em Rural Sociology}, vol.~85, no.~3, pp.~589--622, 2020.

\bibitem{haffajee2019characteristics}
R.~L. Haffajee, L.~A. Lin, A.~S. Bohnert, and J.~E. Goldstick, ``Characteristics of {US} counties with high opioid overdose mortality and low capacity to deliver medications for opioid use disorder,'' {\em JAMA network open}, vol.~2, no.~6, pp.~e196373--e196373, 2019.

\bibitem{liu2023geo}
M.~Liu, J.~M. Caplan, L.~W. Kennedy, I.~K. Moise, D.~J. Feaster, V.~E. Horigian, J.~M. Roll, S.~M. McPherson, and J.~S. Rao, ``Geo-spatial risk factor analysis for drug overdose death in south florida from 2014 to 2019, and the independent contribution of social determinants of health,'' {\em Drug and alcohol dependence}, vol.~248, p.~109931, 2023.

\bibitem{khanal2021role}
B.~Khanal, ``Role of social connectedness in response to a public health crisis: The case study of the flint water crisis,'' {\em Available at SSRN 3969415}, 2021.

\bibitem{charoenwong2020social}
B.~Charoenwong, A.~Kwan, and V.~Pursiainen, ``Social connections with {COVID}-19--affected areas increase compliance with mobility restrictions,'' {\em Science Advances}, vol.~6, no.~47, p.~eabc3054, 2020.

\bibitem{holtz2020interdependence}
D.~Holtz, M.~Zhao, S.~G. Benzell, C.~Y. Cao, M.~A. Rahimian, J.~Yang, J.~Allen, A.~Collis, A.~Moehring, T.~Sowrirajan, {\em et~al.}, ``Interdependence and the cost of uncoordinated responses to {COVID-19},'' {\em Proceedings of the National Academy of Sciences}, vol.~117, no.~33, pp.~19837--19843, 2020.

\bibitem{bailey2024social}
M.~Bailey, D.~Johnston, M.~Koenen, T.~Kuchler, D.~Russel, and J.~Stroebel, ``Social networks shape beliefs and behavior: Evidence from social distancing during the {COVID-19} pandemic,'' {\em Journal of Political Economy Microeconomics}, vol.~2, no.~3, pp.~000--000, 2024.

\bibitem{basu2023social}
A.~K. Basu, N.~H. Chau, and O.~Firsin, ``Social connections and {COVID-19} vaccination,'' in {\em Discussion Paper Series}, Institute of Labor Economics (IZA), 2023.

\bibitem{vahedi2021spatiotemporal}
B.~Vahedi, M.~Karimzadeh, and H.~Zoraghein, ``Spatiotemporal prediction of {COVID-19} cases using inter-and intra-county proxies of human interactions,'' {\em Nature Communications}, vol.~12, no.~1, p.~6440, 2021.

\bibitem{chetty2022social-I}
R.~Chetty, M.~O. Jackson, T.~Kuchler, J.~Stroebel, N.~Hendren, R.~B. Fluegge, S.~Gong, F.~Gonzalez, A.~Grondin, M.~Jacob, {\em et~al.}, ``Social capital ii: determinants of economic connectedness,'' {\em Nature}, pp.~1--13, 2022.

\bibitem{chetty2022social-II}
R.~Chetty, M.~O. Jackson, T.~Kuchler, J.~Stroebel, N.~Hendren, R.~B. Fluegge, S.~Gong, F.~Gonzalez, A.~Grondin, M.~Jacob, {\em et~al.}, ``Social capital {I}: measurement and associations with economic mobility,'' {\em Nature}, vol.~608, no.~7921, pp.~108--121, 2022.

\bibitem{cutter2021social}
C.~M. Cutter, R.~C. Larson, and M.~Abir, ``Social network theory—an underutilized opportunity to align innovative methods with the demands of the opioid epidemic,'' {\em The American Journal of Drug and Alcohol Abuse}, vol.~47, no.~3, pp.~305--310, 2021.

\bibitem{dasgupta2018opioid}
N.~Dasgupta, L.~Beletsky, and D.~Ciccarone, ``Opioid crisis: no easy fix to its social and economic determinants,'' {\em American Journal of Public Health}, vol.~108, no.~2, pp.~182--186, 2018.

\bibitem{friedman2010regularization}
J.~Friedman, T.~Hastie, and R.~Tibshirani, ``Regularization paths for generalized linear models via coordinate descent,'' {\em Journal of Statistical Software}, vol.~33, no.~1, p.~1, 2010.

\bibitem{lee2010efficient}
L.-f. Lee and X.~Liu, ``Efficient {GMM} estimation of high order spatial autoregressive models with autoregressive disturbances,'' {\em Econometric Theory}, vol.~26, no.~1, pp.~187--230, 2010.

\end{thebibliography}
\bibliographystyle{ieeetr}

\clearpage

\appendix
\newtagform{supplementary}[S.]()
\renewcommand{\thetable}{S\arabic{table}}
\renewcommand{\thefigure}{S\arabic{figure}}
\renewcommand{\figurename}{\bf Supplementary Figure}
\renewcommand{\tablename}{\bf Supplementary Table}
\renewcommand{\thepage}{S\arabic{page}}
\renewcommand{\thesection}{S\arabic{section}}
\setcounter{figure}{0}
\setcounter{table}{0}
\setcounter{page}{1}
\usetagform{supplementary}
\noindent {\Huge \bf Supplementary Information}\\


{\tableofcontents}

\section{Additional related work}\label{sec:lirev}

SCI data has been used in different social settings. For example, in counties where people have more concentrated social networks within the counties than outside, they also tend to have worse socioeconomic outcomes. Specifically, these counties exhibit lower average income, lower levels of education, higher rates of teenage births, lower life expectancy, less social capital, and reduced social mobility \cite{bailey2018social}. Diemer and Regan (2021) use SCI data to establish informal social interaction and highlight where social learning and knowledge flow might occur independently of geographical constraints \cite{diemer2022no}. In public health applications, Binod Khanal (2021) uses SCI to explain the variability of bottled water as avoidance behavior in the United States during the Flint water crisis \cite{khanal2021role}. Charonweng et al. (2020) use SCI to show how social networks are conduits of information about the pandemic and an economically important factor that affects compliance with and impact of mobility restrictions \cite{charoenwong2020social}. The findings of their work also suggest that SCI is likely to be a good proxy for real-life social connections. SCI data has also been used to understand better public health policies, for example, by studying the social network spillovers of regional lockdown policies during the pandemic \cite{holtz2020interdependence}. Others have used SCI data to show the effect of social connections on beliefs and behaviors toward social distancing \cite{bailey2024social} and vaccination \cite{basu2023social}. The results of Basu et al. \cite{basu2023social} highlight that increased vaccination rates among Facebook friends correlate with higher vaccination rates in the individual's county. Vahedi et al. (2021) show the utility of SCI in spatio-temporal models to predict hot spots and incidence rates of COVID-19 at the county level in the United States \cite{vahedi2021spatiotemporal}. In all these recent works, SCI is a primary tool for understanding real-life social connections in various domains and acts as an essential tool to represent both economic and social interactions. As indicated in the analysis by Chetty et al.(2022) \cite{chetty2022social-I,chetty2022social-II}, SCI serves as a proxy for friendship links that mediate economic mobility \cite{chetty2022social-I} and affect other outcomes ranging from education to health \cite{chetty2022social-II}. Following this growing literature, we use SCI as a proxy for real-life social connections and not as merely a metric of online connections. Through our analysis, we argue that the usage of SCI is not limited to economic outcomes and can be used to model intricate human networks that underlie public health crises such as the opioid epidemic. 

\section{Social and spatial proximity at the state level and across counties in PA} \label{sec:county maps}

Figures \ref{fig:state-social-network-diagram} and \ref{fig:State-Spatial-spread-of-primary-variable} display the state-level dispersion of the proximity weights and variables of interest.  We formally define the aggregated state-level social and spatial proximity weights as follows: \begin{align*}
w_{xz} = \frac{ \sum_{i \in x} \sum_{j \in z} w_{ij} }{ \sum_{ z' \neq x } \sum_{i \in x} \sum_{j \in z'} w_{ij} }, \; \mbox{ and } \; 
a_{xz} = \frac{1 + \frac{1}{d_{xz}}}{\sum_{m \neq x} (1 + \frac{1}{d_{xm}})}
, 
\end{align*} where $w_{ij}$ is the social proximity weights (defined in the main text) between county $i$ in state $x$ and county $j$ in state $z$, $z'$ represents the set of states excluding state $x$ and $d_{xz}$ represent the distance between states $x$ and $z$. Similarly, we calculate ``deaths in social proximity" and ``deaths in spatial proximity" at the state level as follows:
\begin{align*}
 s_{-x} = \sum_{z\neq x} w_{xz}y_{z},\; \mbox{ and } \; d_{-i} =\sum_{z\neq x} a_{xz}y_{z},
 \end{align*} where $y_{z}$ is the death rate per $100,000$ in state $z$. It is important to note that our analysis of the effect of ``deaths in social proximity" on overdose deaths is conducted at the county level. Due to the unavailability of state-level Social Connectedness Index (SCI) measures, we aggregated the social proximity weights for visualization of state-level variables in Figures \ref{fig:state-social-network-diagram} and \ref{fig:State-Spatial-spread-of-primary-variable}. Using state-level outcomes in these figures allows us to observe how the aggregated county units contribute to differences across states. This approach also aids in visualizing the network layout, which is otherwise infeasible given the density of the connectivity between counties. 
 
\begin{figure}[!hbt]
   \centering
   \includegraphics[width=\textwidth]{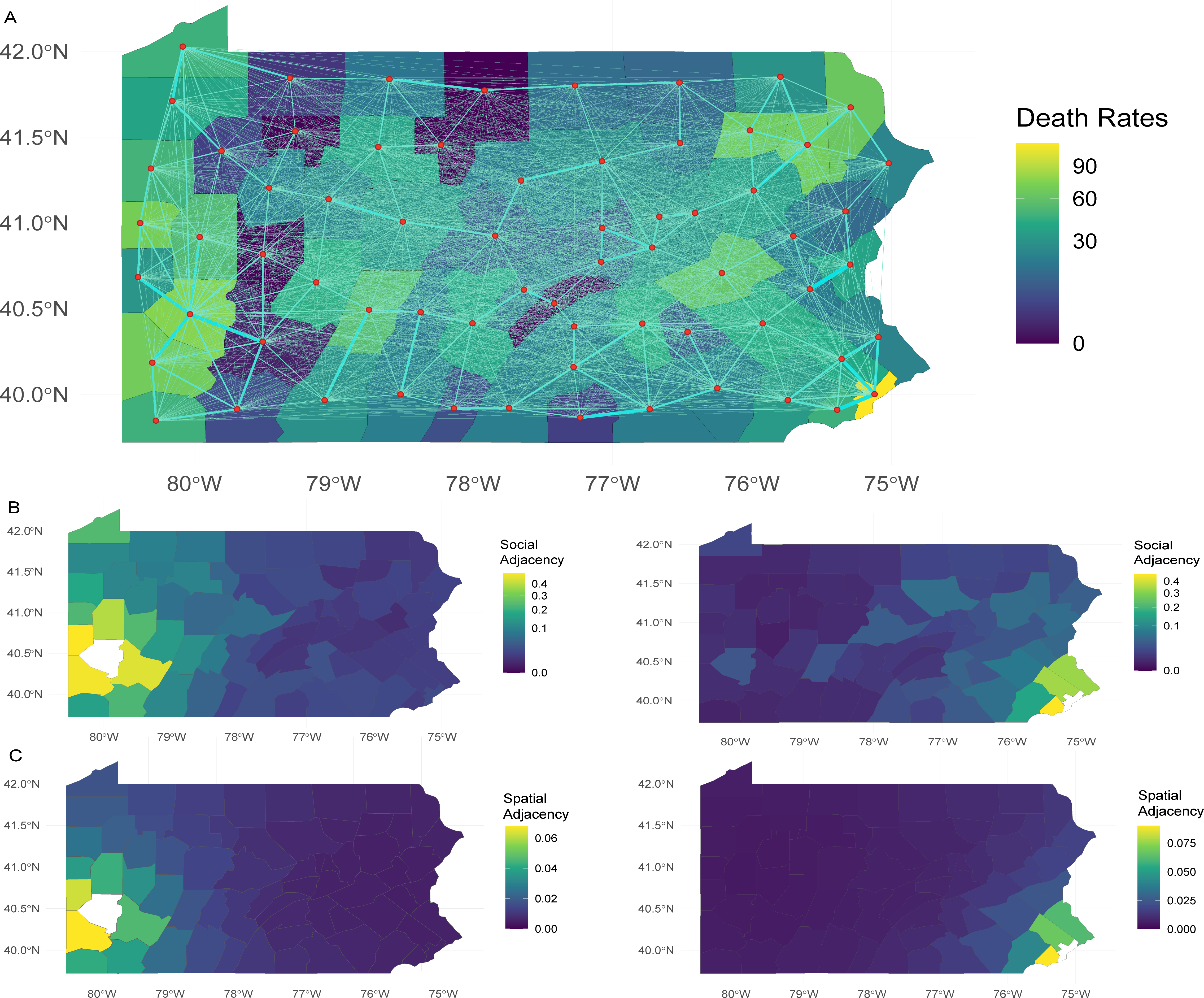}
   \caption{A) The spatial distribution of overdose death rates in 67 counties in Pennsylvania (PA) in 2018 and 2019. The map highlights the higher proportion of opioid overdose death rate in Philadelphia County. Superimposed on this map is a social network diagram with edge widths representing the social proximity weights ($w_{ij}$). B) The two middle maps show the social proximity weights of alter counties to Allegheny County (on the left) and Philadelphia County (on the right). C) The bottom two maps show the spatial proximity weights of alter counties to Allegheny County (on the left) and Philadelphia County (on the right).}
   \label{fig:social-network-diagram}
\end{figure}

\begin{figure}[!hbt]
   \centering
   \includegraphics[width=\textwidth]{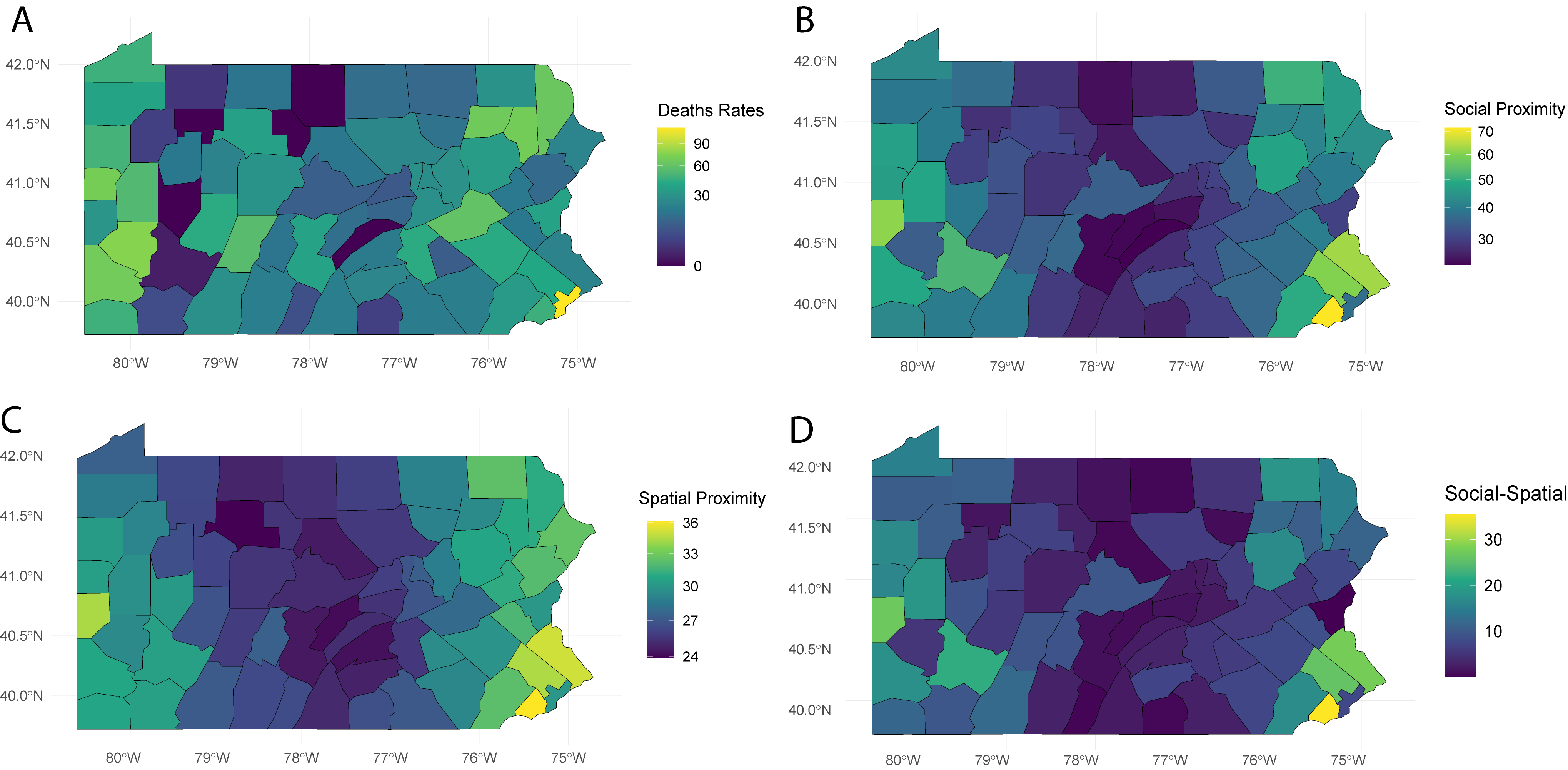}
   \caption{A) The top left map shows the spatial spread of opiod overdose death rates across PA counties. B) The top right map shows the spatial dispersion of death in social proximity ($s_{-i}$) for PA counties. C) The bottom left map shows the geographical spread of ``deaths in spatial proximity" ($d_{-i}$) for PA counties. D) The bottom right map shows the difference between deaths in social and spatial proximity ($s_{-i} - d_{-i}$) in PA counties.}
   \label{fig:Spatial-spread-of-primary-variable}
\end{figure}

 For comparison, we also visualize the layout of social networks and the dispersion of proximity weights between counties in Pennsylvania (PA). Figure \ref{fig:social-network-diagram}A illustrates the social network at the county level measured by their social proximity weights ($w_{ij}$) in PA counties. Figures \ref{fig:social-network-diagram}B and \ref{fig:social-network-diagram}C show the spatial dispersion of the proximity weights for the socially and spatially lagged variables for two ego counties, Allegheny (\ref{fig:social-network-diagram}B) and Philadelphia (\ref{fig:social-network-diagram}C). In Figure \ref{fig:Spatial-spread-of-primary-variable}, we show the spatial distributions of our primary variables of interest: the OOD rates, $s_{-i}$, $d_{-i}$, and their differences ($s_{-i}-d_{-i}$), in PA
counties. 

\section{Covariate selection}
\label{sec:variable-selection}
The opioid epidemic is a complex interplay between social networks and socioeconomic status. Studies have shown that prescription opioids became a means of establishing and maintaining social capital within social networks in low-socioeconomic-status communities \cite{cutter2021social}. Similarly, Gupta et al. (2018) \cite{dasgupta2018opioid} underscore the relationship between the opioid epidemic and its association with socioeconomic factors. To control the influence of socioeconomic factors on ``deaths in social proximity", we used data on social determinants of health (SDOH) available from the Agency for Healthcare Research and Quality (AHRQ). This data set includes information on the physical and health infrastructure and socioeconomic indicators for each county in the United States. We further select the most relevant SDOH covariates using the Least Absolute Shrinkage and Selection Operator (LASSO) to avoid overfitting and multicollinearity in our model. To incorporate domain knowledge variables informed by literature, we specifically use the partially penalized LASSO regression \cite{friedman2010regularization}.

The LASSO objective function for the model we utilize for the analysis:

\begin{align*}
\underset{\beta}{\text{minimize}}\left[\frac{1}{2n} \sum_{i=1}^{n}(y_i - \sum_{j=1}^{p}x_{ij}\beta_j)^2 + \lambda\sum_{j=1}^{p}\pi_j|\beta_j|\right]
\end{align*}

Here $\beta$ is the vector of regression coefficients, $x$ is the design matrix containing SDOH variables, $\pi$ is an indicator variable for the penalized factor, if $\pi_j=0$ then the corresponding covariates are not reduced to zero ensuring that they stay in the model and if $\pi_j=1$ their corresponding covariates are subject to shrinkage, enabling variable selection. We use this LASSO framework to ensure that covariates informed by domain knowledge remain in the regression model, while the LASSO selection operates on other SDOH covariates. Specifically, variables such as the Opioid Dispensing Rate (ODR), naloxone availability, buprenorphine availability, population density, mental health distress rate, political affiliation, and median household income are considered crucial based on domain expertise and are therefore not penalized. The remaining covariates listed in Supplementary Table \ref{tab:Merged_Variables_Coefficients} are subject to selection through the LASSO method. In the model $y$ is the vector of the response variable ``death rates" aggregated over four years (2013-2017) for counties in Pennsylvania. The data source is from the Commonwealth of Pennsylvania Department of Health, Bureau of Health Statistics and Registries. The data source lacks information on `T codes'; consequently, the LASSO selection for SDOH covariates utilizes overall overdose death rates instead of specifically targeting opioid overdose death rates. Here, $n$ is the number of observations  and $p$ is the number of predictors. Performing selection using a different dataset and outcome variable helps avoid post-selection inference issues, ensuring the confidence interval derived from our models has a valid interpretation. The bias introduced by the $L_1$ penalty term forces some regression coefficients to shrink to zero. The zero coefficients are removed from the model, which serves as a way to perform variable selection. To estimate the optimal value of the tuning parameter $\lambda$, we performed a k-fold cross-validation ($k=6$), using the mean squared error (MSE) performance metric. 

Supplementary Table \ref{tab:Merged_Variables_Coefficients} shows the set of coefficient estimates of SDOH variables after selection of LASSO. The selected variables indicate their significance for the death rates. Thus, this Supplementary Table collectively provides a comprehensive understanding of the influential variables that play a significant role in OODs. To isolate the effect of ``deaths in social proximity" on opioid overdose deaths, we used the variables selected by LASSO regression (see Table \ref{tab:Merged_Variables_Coefficients}) as controls in our model. We excluded state-level data on the count of illicit opioids reported from the variable selection pool because this measure lacks variation within Pennsylvania counties; consequently, LASSO would automatically drop this variable even with penalized regression. In Supplementary Table \ref{tab:Merged_Variables_Coefficients}, a ``–" indicates variables that were not selected by the LASSO regression. When analyzing the National Center for Health Statistics (NCHS) data to estimate the effect of "deaths in social proximity" on OODs, we included only the covariates selected by the LASSO, along with state-level control. It is important to note that in our analysis, counties lacking reported values for clinical covariates, specifically the availability of naloxone and buprenorphine, are assumed to have zero values. This assumption is based on the observation that these counties recorded negligible overdose deaths during the years 2018-2019 and have very small populations. In the next section, we discuss the specifics of the robustness checks conducted in our study.

\begin{table}[!htbp]
\centering
\caption{Columns one to four list SDOH variable names, their descriptions, LASSO model coefficients (- values are dropped post-LASSO), and the corresponding covariates in our models.}
\label{tab:Merged_Variables_Coefficients}
\tiny
\begin{tabular}{@{\extracolsep{4pt}}clp{0.2\textwidth}cc} %
\\[-1.8ex]\hline
\hline \\[-1.8ex]
 & \textbf{Variable} & \textbf{Description} & \textbf{Coefficient} & \textbf{Covariates} \\
\hline \\[-1.8ex]
1 & ACS\_TOT\_POP\_POV  & Total population for whom poverty status is determined& - & poverty status \\
2 & AMFAR\_MHFAC\_RATE & Total number of facilities that provide mental health services per 1,000 population & - & mental health service rate \\
3 & ACS\_PCT\_HU\_NO\_VEH & Percentage of housing units with no vehicle available & - & percentage no vehicle \\
4 & ACS\_PCT\_LT\_HS & Percentage of population with less than high school education (ages 25 and over) & - & education level (below high school) \\
5 & POS\_MEAN\_DIST\_ALC & Mean distance to nearest hospital with alcohol/drug abuse care, using pop-weighted centroids in county & $-7.518093 \times 10^{-5}$ & mean distance to hospital \\
6 & ACS\_PCT\_OTHER\_INS & Percentage with other health insurance coverage & - & health insurance coverage \\
7 & CCBP\_TOT\_BWLSTORES\_Rate & Number of beer, wine, liquor stores per 1,000 population & - & alcohol stores per 1,000 Pop \\
8 & AHRF\_TOT\_COM\_HEALTH\_GRANT & Total community health centers, grantees only & - & total community health centers \\
9 & ACS\_PCT\_UNEMPLOY & percentage unemployed (civilian labor force, ages 16+) & $6.249815 \times 10^{-5}$ & percentage unemployed \\
10 & ACS\_MEDIAN\_HH\_INC & median household income & $8.190151 \times 10^{-5}$ & median household income \\
11 & ACS\_MEDIAN\_AGE & Median age of population & - & median age \\
12 & ACS\_PCT\_WHITE & Percentage of population white & - & percentage White \\
13 & ACS\_PCT\_BLACK & Percentage of population black & - & percentage Black \\
14 & ACS\_PCT\_AIAN & Percentage American Indian and Alaska Native & $-2.546380 \times 10^{-5}$ & percentage AIAN \\
15 & ACS\_PCT\_NHPI & Percentage Native Hawaiian and Other Pacific Islander & $1.307321 \times 10^{-5}$ & percentage NHPI \\
16 & ACS\_PCT\_MULT\_Race & Percentage two or more races & - & percentage multi-race \\
17 & ACS\_PCT\_ASIAN & Percentage of population Asian & - & percentage Asian \\
18 & ODR & Opioid Dispensing Rate  & $1.993769 \times 10^{-4}$ & ODR \\
19 & Naloxone\_Available & Availability of naloxone & $7.221718 \times 10^{-4}$ & naloxone available \\
20 & buprenorphine\_Available & Availability of buprenorphine  & $6.462634 \times 10^{-3}$ & buprenorphine available \\
21 & population\_density & Population per square kilometers & $-1.385327 \times 10^{-8}$ & population density \\
22 & frequent\_mental\_health\_distress & Rate of frequent mental health distress & $3.000371 \times 10^{-3}$ & mental health distress rate \\
23 & political\_affiliation & political affiliation measure & $2.921279 \times 10^{-4}$ & political affiliation \\
\hline \\[-1.8ex]
\hline \\[-1.8ex]
\end{tabular}
\end{table}

\section{Robustness checks}
 As robustness checks, we set up linear regression, spatial and network autocorrelation, two-way fixed-effect models, two-stage least squares estimates. These steps help us ensure consistent reporting and further bolster our claims regarding the significance of our socially lagged variable.

\subsection{Linear regression with cluster-robust standard error correction}
\label{sec:section_linear_regression}
As a first robustness check model, we utilize a simple linear regression as shown in the  equation:
\begin{align*} 
    y_i = &  \beta_0 + \beta_1 s_{-i} + \beta_2 d_{-i} + \overline{\beta_3}^T \overline{C}_i+ \overline{\beta_4}^T \overline{X}_i + \epsilon_i
\end{align*}

 Here, $\beta_1$ is the coefficient for ``deaths in social proximity", $\beta_2$ is the coefficient for ``deaths in spatial proximity", $\overline{\beta_3}^T$ denotes the vector of the coefficients associated with clinical covariates, and $\overline{\beta_4}^T$ is the vector of the coefficient associated with SDOH variables. The model allows us to estimate the effect of our socially lagged variable on death rates.
  
To ensure that the inference is reflecting the population residing in the counties across the US, we weight the residuals by population size. Weighing the residuals also enhances the ability of the model to capture the heterogeneity between different spatial units. In our simple linear model, we ensure that we account for the clustering of standard errors using cluster-robust standard error correction. 

Supplementary Tables \ref{tab:simple_linear_model_with_robust_std_clustered_error-eastern-united-state}, \ref{tab:simple_linear_model_with_robust_std_clustered_error-western-united-states}, and \ref{tab:simple_linear_model_with_robust_std_clustered_error-contiguous-united-states} illustrate the results of the linear regression model for the eastern, western-central, and contiguous United States. The socially lagged variable remains significant. However, we lose the statistical significance for ``deaths in spatial proximity" when the geographical focus shifts from counties in the eastern US to counties in the contiguous US. Our previous discussion has highlighted the underlying spatial and network structure of our variable of interest. Spatial and social variables are often susceptible to correlated error terms because of their autocorrelated construction. Hence, as a mitigating measure, we use the spatial error model to account for network and spatial autocorrelation in the model. In the next section, we discuss the network autocorrelation model.
\begin{table}[!htbp] \centering 
  \caption{\footnotesize{Linear regression for measuring the significance of the effect size of ``deaths in social proximity", for counties in the eastern United States}}  
 \label{tab:simple_linear_model_with_robust_std_clustered_error-eastern-united-state} 
\tiny
\begin{tabular}{@{\extracolsep{5pt}}lcc} 
\\[-1.8ex]\hline 
\hline \\[-1.8ex] 
 & \multicolumn{2}{c}{\textit{Dependent variable: death rate per 100,000 people}} \\ 
\cline{2-3} 
\\[-1.8ex] & \textit{OLS} & \textit{Cluster-Robust OLS} \\ 
\\[-1.8ex] & (1) & (2)\\ 
\hline \\[-1.8ex] 
deaths social proximity & 9.187$^{***}$ & 9.187$^{***}$ \\ 
 & (0.978) & (2.414) \\ 
 & & \\ 
deaths spatial proximity & 5.462$^{***}$ & 5.462$^{***}$ \\ 
 & (0.921) & (2.021) \\ 
 & & \\ 
ODR & 54.106$^{***}$ & 54.106$^{**}$ \\ 
 & (12.571) & (23.513) \\ 
 & & \\ 
naloxone available & 162.003$^{***}$ & 162.003$^{*}$ \\ 
 & (17.037) & (85.602) \\ 
 & & \\ 
buprenorphine available & 42.685$^{***}$ & 42.685 \\ 
 & (13.212) & (40.431) \\ 
 & & \\ 
state count illicit opioid reported & 6,303.967$^{**}$ & 6,303.967 \\ 
 & (3,193.753) & (8,982.550) \\ 
 & & \\ 
population density & $-$0.001$^{***}$ & $-$0.001$^{*}$ \\ 
 & (0.0002) & (0.001) \\ 
 & & \\ 
mental health distress rate & 182.710$^{**}$ & 182.710 \\ 
 & (77.001) & (185.534) \\ 
 & & \\ 
political affiliation & $-$3.649$^{***}$ & $-$3.649 \\ 
 & (1.340) & (2.706) \\ 
 & & \\ 
percentage unemployed & 68.805$^{***}$ & 68.805$^{**}$ \\ 
 & (9.677) & (29.286) \\ 
 & & \\ 
mean distance to hospital & $-$72.700$^{***}$ & $-$72.700$^{***}$ \\ 
 & (7.093) & (11.279) \\ 
 & & \\ 
median household income & $-$24.389$^{***}$ & $-$24.389 \\ 
 & (6.898) & (15.699) \\ 
 & & \\ 
percentage AIAN & $-$41.111 & $-$41.111$^{*}$ \\ 
 & (35.741) & (23.164) \\ 
 & & \\ 
percentage NHPI & $-$15.308 & $-$15.308 \\ 
 & (12.245) & (13.041) \\ 
 & & \\ 
constant & 2.770 & 2.770 \\ 
 & (10.378) & (26.515) \\ 
 & & \\ 
\hline \\[-1.8ex] 
observations  & 1,606 &  \\ 
R$^{2}$ & 0.464 &  \\ 
adjusted R$^{2}$ & 0.460 &  \\ 
residual std. error & 3.179 (df = 1590) &  \\ 
F statistic & 94.409$^{***}$ (df = 14; 1590) &  \\ 
\hline 
\hline \\[-1.8ex] 
\textit{Note:}  & \multicolumn{2}{r}{$^{*}$p$<$0.1; $^{**}$p$<$0.05; $^{***}$p$<$0.01} \\ 
\end{tabular} 
\end{table} 

\begin{table}[!htbp] \centering 
  \caption{\footnotesize{Linear regression for measuring the significance of the effect size of ``deaths in social proximity", for counties in the western and central United States}}  
 \label{tab:simple_linear_model_with_robust_std_clustered_error-western-united-states} 
\tiny
\begin{tabular}{@{\extracolsep{5pt}}lcc} 
\\[-1.8ex]\hline 
\hline \\[-1.8ex] 
 & \multicolumn{2}{c}{\textit{Dependent variable: death rate per 100,000 people}} \\ 
\cline{2-3} 
\\[-1.8ex] & \textit{OLS} & \textit{Cluster-Robust OLS} \\ 
\\[-1.8ex] & (1) & (2)\\ 
\hline \\[-1.8ex] 
deaths social proximity & 5.768$^{***}$ & 5.768$^{***}$ \\ 
 & (0.340) & (0.637) \\ 
 & & \\ 
deaths spatial proximity & 0.128 & 0.128 \\ 
 & (0.528) & (1.873) \\ 
 & & \\ 
ODR & 17.101$^{***}$ & 17.101$^{**}$ \\ 
 & (3.847) & (8.698) \\ 
 & & \\ 
naloxone available & 69.584$^{***}$ & 69.584$^{**}$ \\ 
 & (10.675) & (30.149) \\ 
 & & \\ 
buprenorphine available & 43.633$^{***}$ & 43.633$^{**}$ \\ 
 & (8.586) & (19.757) \\ 
 & & \\ 
state count illicit opioid reported & 6,468.861 & 6,468.861 \\ 
 & (4,288.239) & (9,546.581) \\ 
 & & \\ 
population density & 0.006$^{***}$ & 0.006$^{***}$ \\ 
 & (0.0005) & (0.001) \\ 
 & & \\ 
mental health distress rate & 58.461 & 58.461 \\ 
 & (38.647) & (147.098) \\ 
 & & \\ 
political affiliation & 0.578 & 0.578 \\ 
 & (0.723) & (2.070) \\ 
 & & \\ 
percentage unemployed & $-$6.495 & $-$6.495 \\ 
 & (6.179) & (9.489) \\ 
 & & \\ 
mean distance to hospital & $-$15.183$^{***}$ & $-$15.183 \\ 
 & (3.604) & (10.792) \\ 
 & & \\ 
median household income & $-$4.777 & $-$4.777 \\ 
 & (3.467) & (10.669) \\ 
 & & \\ 
percentage AIAN & $-$7.516 & $-$7.516 \\ 
 & (6.795) & (12.271) \\ 
 & & \\ 
percentage NHPI & $-$2.114 & $-$2.114 \\ 
 & (3.862) & (6.128) \\ 
 & & \\ 
constant & 0.926 & 0.926 \\ 
 & (5.349) & (22.011) \\ 
 & & \\ 
\hline \\[-1.8ex] 
observations  & 1,502 &  \\ 
R$^{2}$ & 0.459 &  \\ 
adjusted R$^{2}$ & 0.454 &  \\ 
residual std. error & 1.306 (df = 1486) &  \\ 
F statistic & 90.049$^{***}$ (df = 14; 1486) &  \\ 
\hline 
\hline \\[-1.8ex] 
\textit{Note:}  & \multicolumn{2}{r}{$^{*}$p$<$0.1; $^{**}$p$<$0.05; $^{***}$p$<$0.01} \\ 
\end{tabular} 
\end{table} 

\begin{table}[!htbp] \centering 
  \caption{\footnotesize{Linear regression for measuring the significance of the effect size of ``deaths in social proximity", for counties in the contiguous United States.}}
 \label{tab:simple_linear_model_with_robust_std_clustered_error-contiguous-united-states} 
\tiny
\begin{tabular}{@{\extracolsep{5pt}}lcc} 
\\[-1.8ex]\hline 
\hline \\[-1.8ex] 
 & \multicolumn{2}{c}{\textit{Dependent variable: death rate per 100,000 people}} \\ 
\cline{2-3} 
\cline{2-3} 
\\[-1.8ex] & \textit{OLS} & \textit{Cluster-Robust OLS} \\ 
\\[-1.8ex] & (1) & (2)\\ 
\hline \\[-1.8ex] 
deaths social proximity & 12.503$^{***}$ & 12.503$^{***}$ \\ 
 & (0.706) & (2.163) \\ 
 & & \\ 
deaths spatial proximity & 1.132$^{*}$ & 1.132 \\ 
 & (0.648) & (1.526) \\ 
 & & \\ 
ODR & 49.014$^{***}$ & 49.014$^{***}$ \\ 
 & (8.187) & (18.479) \\ 
 & & \\ 
naloxone available & 153.300$^{***}$ & 153.300$^{***}$ \\ 
 & (11.252) & (57.614) \\ 
 & & \\ 
buprenorphine available & 56.048$^{***}$ & 56.048 \\ 
 & (10.038) & (41.411) \\ 
 & & \\ 
state count illicit opioid reported & 11,396.620$^{***}$ & 11,396.620 \\ 
 & (2,391.373) & (8,899.337) \\ 
 & & \\ 
population density & 0.0003$^{*}$ & 0.0003 \\ 
 & (0.0002) & (0.001) \\ 
 & & \\ 
mental health distress rate & 21.961 & 21.961 \\ 
 & (45.829) & (119.935) \\ 
 & & \\ 
political affiliation & $-$3.589$^{***}$ & $-$3.589$^{*}$ \\ 
 & (0.829) & (1.893) \\ 
 & & \\ 
percentage unemployed & 40.118$^{***}$ & 40.118$^{*}$ \\ 
 & (6.755) & (23.199) \\ 
 & & \\ 
mean distance to hospital & $-$50.508$^{***}$ & $-$50.508$^{***}$ \\ 
 & (5.560) & (12.930) \\ 
 & & \\ 
median household income & $-$17.588$^{***}$ & $-$17.588 \\ 
 & (4.330) & (10.696) \\ 
 & & \\ 
percentage AIAN & $-$17.954 & $-$17.954 \\ 
 & (10.987) & (13.417) \\ 
 & & \\ 
percentage NHPI & $-$9.727$^{*}$ & $-$9.727 \\ 
 & (5.689) & (10.393) \\ 
 & & \\ 
constant & 9.849 & 9.849 \\ 
 & (6.194) & (16.294) \\ 
 & & \\ 
\hline \\[-1.8ex] 
observations & 3,108 &  \\ 
R$^{2}$ & 0.502 &  \\ 
adjusted R$^{2}$ & 0.500 &  \\ 
residual std. error & 1.899 (df = 3092) &  \\ 
F Statistic & 222.608$^{***}$ (df = 14; 3092) &  \\ \\ 
\hline 
\hline \\[-1.8ex] 
\textit{Note:}  & \multicolumn{2}{r}{$^{*}$p$<$0.1; $^{**}$p$<$0.05; $^{***}$p$<$0.01} \\ 
\end{tabular} 
\end{table} 
 
\clearpage
\subsection{Network autocorrelation}
\label{sec:network_autocorrelation}
Autocorrelation, in essence, refers to the tendency of a variable to be correlated with itself at different points in space or time. In the context of our study, we anticipate encountering correlated error terms when analyzing death rates related to social proximity. This likelihood comes from the inherent structural characteristics and the defining attributes of our variable of interest. Thus, we utilize the Spatial Error Model (SEM) to account for spatial and network autocorrelation. SEM is a specific linear regression model that accounts for spatial dependence in error terms. The so-called network autocorrelation model for our study is defined by, 
$
y_i =  \gamma_0 + \gamma_1 s_{-i} + \gamma_2 d_{-i} + \overline{\gamma_3}^T \overline{C}_i + \overline{\gamma_4}^T \overline{X}_i + u_i   
$
, where $\gamma_1$ represents the coefficient associated with ``deaths in social proximity", $\gamma_2$ illustrates the coefficient size related to ``deaths in spatial proximity", $\overline{\gamma_3}^T$ denotes the vector of the coefficient associated with clinical covariates, and $\overline{\gamma_4}^T$ be the vector of the coefficient associated with SDOH variables. The error term $u$ is assumed to be spatially autocorrelated, which means that geographically close observations  are more likely to have similar error terms than distant observations . 

To account for autocorrelation in error terms, the SEM assumes that error terms can be decomposed into two components: an autocorrelated error term \(u\) and a noise error term \(\overline{\epsilon}\), where \(\overline{\epsilon} \sim N(0, \sigma^2I)\). The autocorrelated error term is modeled as a linear combination of the error terms of neighboring observations , weighted by a spatial weight matrix \(W\). The error model is given by:
\begin{align*}    
u_i =  \lambda \sum_j w_{ij} u_j + \epsilon_i.
\end{align*}

In the weight matrix $W$, each element 
\begin{align*}
 w_{ij} &= \frac{n_{j}\mbox{SCI}_{ij}}{\sum_{k\neq i} n_{k} \mbox{SCI}_{ik}},
\end{align*} where $\lambda$ is a parameter that captures the strength of network autocorrelation, and $W$ calculates the weight for the connection between two areas i and j in a network. The noise in error term $\epsilon$ is assumed to be independently and identically distributed (IID) with mean zero and variance $\sigma^2$. To estimate the model, we utilize the \texttt{spatialreg} package in R, specifically employing the \texttt{errorsarlm} function. The function implements maximum-likelihood estimation for spatial simultaneous autoregressive error models. The parameter $\lambda$ is initially estimated using the \texttt{optimize} function. Subsequently, the $\beta$ and other parameters are estimated using generalized least squares. Same estimation method is used when accounting for spatial autocorrelation model.

The findings of the network autocorrelation model are shown in the Supplementary Tables \ref{tab:network_spatial_autocorrelation_model_for_eastern_united_States}, \ref{tab:network_spatial_autocorrelation_model_for_western_united_States} and \ref{tab:network_spatial_autocorrelation_model_for_entire_united_States} for the eastern, western-central and contiguous United States. To further solidify our analysis, we account for spatial dependence employing spatial autocorrelation.

\subsection{Spatial autocorrelation}
\label{sec:spatial_autocorrelation}
From a methodological perspective, the network and spatial autocorrelation share many similarities. However, a key difference is induced through the definition of weights. The weights are designed to encapsulate the underlying network structure in a network autocorrelation model. In contrast, in the case of spatial autocorrelation, the weights are formulated to capture the geographical dependencies.

The error model is defined as follows, where $a_{ij}$ is the spatial weights,
\begin{align*} 
u_i    &=  \lambda \sum_j a_{ij} u_j + \epsilon_i\, ,\\
a_{ij} &= \frac{1 + \frac{1}{d_{ij}}}{\sum_{k \neq i} (1 + \frac{1}{d_{ik}})}\, .
\end{align*}

It is important to note that the equations are used similarly for both the network and the spatial autocorrelation models. However, within the scope of this study, the only distinction between the two models is the definition of these weights. Supplementary Tables \ref{tab:network_spatial_autocorrelation_model_for_eastern_united_States}, \ref{tab:network_spatial_autocorrelation_model_for_western_united_States} and \ref{tab:network_spatial_autocorrelation_model_for_entire_united_States} show the results of the autocorrelation model for the eastern, western-central, and contiguous United States. This result reinforces the validity of our findings derived from simple linear regression and supports the significance obtained for $s_{-i}$. The models we have considered thus far account for heterogeneity across spatial boundaries but do not fully capture variations across space and time. Therefore, as a final robustness check, we will incorporate a two-way fixed effect model into our analysis to better capture these nuances.

\begin{table}[!htbp] \centering 
  \caption{\footnotesize{Autocorrelation Models for the counties in the eastern United States}}
\label{tab:network_spatial_autocorrelation_model_for_eastern_united_States} 
\tiny
\begin{tabular}{@{\extracolsep{5pt}}lcc} 
\\[-1.8ex]\hline 
\hline \\[-1.8ex] 
 & \multicolumn{2}{c}{\textit{Dependent variable: death rate per 100,000 people}} \\ 
\cline{2-3} \\
\\[-1.8ex] & \textit{Network} & \textit{Spatial} \\ 
\\[-1.8ex] & (1) & (2)\\
\hline \\[-1.8ex] 
deaths social proximity & 18.791$^{***}$ & 13.314$^{***}$ \\ 
 & (0.644) & (0.894) \\ 
 & & \\ 
deaths spatial proximity & $-$2.335$^{***}$ & 1.296 \\ 
 & (0.624) & (1.035) \\ 
 & & \\ 
ODR & 47.971$^{***}$ & 53.261$^{***}$ \\ 
 & (6.504) & (7.209) \\ 
 & & \\ 
naloxone available & 31.139$^{***}$ & 31.797$^{***}$ \\ 
 & (10.486) & (11.219) \\ 
 & & \\ 
buprenorphine available & 7.028 & 17.375$^{***}$ \\ 
 & (5.426) & (6.631) \\ 
 & & \\ 
state count illicit opioid reported & $-$1,945.999 & $-$130.128 \\ 
 & (1,756.263) & (2,477.706) \\ 
 & & \\ 
population density & 0.001$^{**}$ & 0.002$^{**}$ \\ 
 & (0.001) & (0.001) \\ 
 & & \\ 
mental health distress rate & $-$160.796$^{***}$ & $-$89.597 \\ 
 & (42.584) & (56.496) \\ 
 & & \\ 
political affiliation & $-$6.124$^{***}$ & $-$4.957$^{***}$ \\ 
 & (1.136) & (1.330) \\ 
 & & \\ 
percentage unemployed & 7.073 & 6.073 \\ 
 & (4.791) & (5.503) \\ 
 & & \\ 
mean distance to hospital & $-$19.737$^{***}$ & $-$26.527$^{***}$ \\ 
 & (3.362) & (3.983) \\ 
 & & \\ 
median household income & $-$22.393$^{***}$ & $-$27.317$^{***}$ \\ 
 & (5.596) & (7.087) \\ 
 & & \\ 
percentage AIAN & 47.620$^{***}$ & 45.101$^{***}$ \\ 
 & (13.005) & (14.542) \\ 
 & & \\ 
percentage NHPI & 1.963 & 4.928 \\ 
 & (6.104) & (6.411) \\ 
 & & \\ 
constant & 53.460$^{***}$ & 43.923$^{***}$ \\ 
 & (6.393) & (8.606) \\ 
 & & \\ 
\hline \\[-1.8ex] 
observations  & 1,606 & 1,606 \\ 
log likelihood & $-$6,931.226 & $-$6,944.193 \\ 
$\sigma^{2}$ & 322.701 & 333.249 \\ 
$\lambda$ & $-$0.81195 & 0.68115 \\
Akaike Inf. Crit. & 13,896.450 & 13,922.390 \\ 
Wald test (df = 1) & 65.324$^{***}$ & 11.167$^{***}$ \\ 
LR Test (df = 1) & 28.489$^{***}$ & 2.554 \\ 
\hline 
\hline \\[-1.8ex] 
\textit{Note:}  & \multicolumn{2}{r}{$^{*}$p$<$0.1; $^{**}$p$<$0.05; $^{***}$p$<$0.01} \\ 
\end{tabular} 
\end{table} 

\begin{table}[!htbp] \centering 
  \caption{\footnotesize{Autocorrelation models for the counties in the western and central United States}}
  \label{tab:network_spatial_autocorrelation_model_for_western_united_States} 
  \tiny
\begin{tabular}{@{\extracolsep{5pt}}lcc} 
\\[-1.8ex]\hline 
\hline \\[-1.8ex] 
 & \multicolumn{2}{c}{\textit{Dependent variable:}} \\ 
\cline{2-3} 
\\[-1.8ex] & \multicolumn{2}{c}{death rates} \\ 
\\[-1.8ex] & \textit{Network} & \textit{Spatial} \\ 
\\[-1.8ex] & (1) & (2)\\
\hline \\[-1.8ex] 
deaths social proximity & 7.284$^{***}$ & 6.454$^{***}$ \\ 
 & (0.405) & (0.566) \\ 
 & & \\ 
deaths spatial proximity & 0.058 & 0.400 \\ 
 & (0.473) & (0.662) \\ 
 & & \\ 
ODR & 6.515$^{***}$ & 6.862$^{***}$ \\ 
 & (2.436) & (2.559) \\ 
 & & \\ 
naloxone available & $-$1.754 & $-$3.460 \\ 
 & (5.690) & (5.905) \\ 
 & & \\ 
buprenorphine available & 24.510$^{***}$ & 30.127$^{***}$ \\ 
 & (5.681) & (6.021) \\ 
 & & \\ 
state count illicit opioid reported & 568.756 & 712.539 \\ 
 & (2,520.982) & (3,707.215) \\ 
 & & \\ 
population density & 0.009$^{***}$ & 0.010$^{***}$ \\ 
 & (0.001) & (0.002) \\ 
 & & \\ 
mental health distress rate & $-$81.327$^{***}$ & $-$54.738$^{*}$ \\ 
 & (23.426) & (30.530) \\ 
 & & \\ 
political affiliation & $-$1.077 & $-$0.716 \\ 
 & (0.980) & (1.093) \\ 
 & & \\ 
percentage unemployed & 6.172$^{*}$ & 5.312 \\ 
 & (3.741) & (4.075) \\ 
 & & \\ 
mean distance to hospital & $-$1.382 & $-$1.576 \\ 
 & (2.355) & (2.587) \\ 
 & & \\ 
median household income & $-$7.234$^{**}$ & $-$5.256 \\ 
 & (3.648) & (4.007) \\ 
 & & \\ 
percentage AIAN & 7.932$^{**}$ & 6.295 \\ 
 & (3.299) & (4.160) \\ 
 & & \\ 
percentage NHPI & $-$0.985 & $-$0.638 \\ 
 & (4.439) & (4.760) \\ 
 & & \\ 
constant & 18.843$^{***}$ & 14.717$^{***}$ \\ 
 & (3.716) & (4.621) \\ 
 & & \\ 
\hline \\[-1.8ex] 
observations  & 1,502 & 1,502 \\ 
log likelihood $-$5,860.207 & $-$5,868.434 \\ 
$\sigma^{2}$ & 142.595 & 144.851 \\ 
$\lambda$ & $-$0.54591 & 0.52449 \\
Akaike Inf. Crit. & 11,754.410 & 11,770.870 \\ 
Wald test (df = 1) & 22.991$^{***}$ & 3.504$^{*}$ \\ 
LR test (df = 1) & 18.083$^{***}$ & 1.629 \\ 
\hline 
\hline 
\hline \\[-1.8ex] 
\textit{Note:}  & \multicolumn{2}{r}{$^{*}$p$<$0.1; $^{**}$p$<$0.05; $^{***}$p$<$0.01} \\ 
\end{tabular} 
\end{table} 

\begin{table}[!htbp] \centering 
  \caption{\footnotesize{Autocorrelation models for counties in the contiguous United States}}
\label{tab:network_spatial_autocorrelation_model_for_entire_united_States} 
\tiny
\begin{tabular}{@{\extracolsep{5pt}}lcc} 
\\[-1.8ex]\hline 
\hline \\[-1.8ex] 
 & \multicolumn{2}{c}{\textit{Dependent variable: death rate per 100,000 people}} \\ 
\cline{2-3} 
\\[-1.8ex] & \textit{Network} & \textit{Spatial} \\ 
\\[-1.8ex] & (1) & (2)\\
\hline \\[-1.8ex] 
deaths social proximity & 17.894$^{***}$ & 14.517$^{***}$ \\ 
 & (0.509) & (0.671) \\ 
 & & \\ 
deaths spatial proximity & $-$3.386$^{***}$ & $-$1.516$^{*}$ \\ 
 & (0.481) & (0.795) \\ 
 & & \\ 
ODR & 36.510$^{***}$ & 39.822$^{***}$ \\ 
 & (4.320) & (4.596) \\ 
 & & \\ 
naloxone available & 21.914$^{***}$ & 19.275$^{***}$ \\ 
 & (6.056) & (6.314) \\ 
 & & \\ 
buprenorphine available & 13.723$^{***}$ & 26.807$^{***}$ \\ 
 & (4.535) & (5.338) \\ 
 & & \\ 
state count illicit opioid reported & 363.577 & 1,052.167 \\ 
 & (1,446.788) & (1,916.814) \\ 
 & & \\ 
population density & 0.003$^{***}$ & 0.003$^{***}$ \\ 
 & (0.0005) & (0.001) \\ 
 & & \\ 
mental health distress rate & $-$124.751$^{***}$ & $-$92.180$^{***}$ \\ 
 & (23.424) & (30.657) \\ 
 & & \\ 
political affiliation & $-$4.311$^{***}$ & $-$4.086$^{***}$ \\ 
 & (0.790) & (0.892) \\ 
 & & \\ 
percentage unemployed & 1.457 & 1.790 \\ 
 & (3.439) & (3.787) \\ 
 & & \\ 
mean distance to hospital & $-$6.814$^{***}$ & $-$9.530$^{***}$ \\ 
 & (2.585) & (2.948) \\ 
 & & \\ 
median household income & $-$12.265$^{***}$ & $-$14.344$^{***}$ \\ 
 & (3.653) & (4.286) \\ 
 & & \\ 
percentage AIAN & 20.064$^{***}$ & 17.316$^{***}$ \\ 
 & (3.815) & (4.736) \\ 
 & & \\ 
percentage NHPI & $-$0.048 & 1.835 \\ 
 & (5.213) & (5.787) \\ 
 & & \\ 
constant & 35.723$^{***}$ & 32.721$^{***}$ \\ 
 & (3.625) & (5.584) \\ 
 & & \\ 
\hline \\[-1.8ex] 
observations  & 3,108 & 3,108 \\ 
log likelihood & $-$13,021.780 & $-$13,025.660 \\ 
$\sigma^{2}$ & 253.641 & 255.359 \\ 
$\lambda$ & $-$0.58906 & 0.90999 \\
Akaike Inf. Crit. & 26,077.560 & 26,085.310 \\ 
Wald test (df = 1) & 47.679$^{***}$ & 214.894$^{***}$ \\ 
LR test (df = 1) & 24.566$^{***}$ & 16.817$^{***}$ \\ 
\hline \\[-1.8ex] 
\textit{Note:}  & \multicolumn{2}{r}{$^{*}$p$<$0.1; $^{**}$p$<$0.05; $^{***}$p$<$0.01} \\ 
\end{tabular} 
\end{table}

\clearpage
\subsection{Two-way fixed-effect model}
\label{sec:two-way-fixed-effect}
The two-way fixed-effects model accounts for cross-sectional heterogeneity between counties in the US and periods (year). It controls for any unobserved space- and time-invariant characteristics that may be correlated with the covariates, furthering more accurate and unbiased estimates. Mathematically, we can express the two-way fixed effect model by 
\begin{align*}
    y_{it} = &\ \zeta_0 + \zeta_1s_{-it} + \zeta_2d_{-it} + \overline{\zeta_3}^T \overline{C}_{it} + \overline{\zeta_4}^T \overline{X}_{it} + \ \mu_i + \phi_t + \epsilon_{it}.
\end{align*}

Here, $\mu_i$ and $\phi_t$ account for spatial and temporal heterogeneity. Supplementary Tables \ref{tab:two-way-fixed-effect-eastern-united-states}, \ref{tab:two-way-fixed-effect-western-united-states} and \ref{tab:two-way-fixed-effect-entire-united-states} show the result obtained from the two-way fixed effect model in the eastern, western-central and the entire contiguous US. After controlling for covariates from SDOH, ``deaths in spatial proximity", and clinical covariates, we still witness statistically significant effect sizes for ``deaths in social proximity."

\begin{table}[!htbp] \centering 
  \caption{\footnotesize{Two-way fixed-effect model for counties in the eastern United States}}
  \label{tab:two-way-fixed-effect-eastern-united-states} 
  \tiny
\begin{tabular}{@{\extracolsep{5pt}}lc} 
\\[-1.8ex]\hline 
\hline \\[-1.8ex] 
 & \multicolumn{1}{c}{\textit{Dependent variable: death rate per 100,000 people }} \\ 
\cline{2-2} \\
\hline \\[-1.8ex] 
deaths social proximity & 3.366$^{***}$ \\ 
 & (0.428) \\ 
 & \\ 
deaths spatial proximity & 3.078$^{***}$ \\ 
 & (0.784) \\ 
 & \\ 
ODR & 16.025 \\ 
 & (32.745) \\ 
 & \\ 
naloxone available & 7,593.117$^{***}$ \\ 
 & (748.879) \\ 
 & \\ 
buprenorphine available & 0.002$^{***}$ \\ 
 & (0.0001) \\ 
 & \\ 
state count illicit opioid reported & $-$1,220.069 \\ 
 & (3,935.194) \\ 
 & \\ 
population density & $-$0.001$^{***}$ \\ 
 & (0.0001) \\ 
 & \\ 
mental health distress rate & 351.109$^{***}$ \\ 
 & (35.178) \\ 
 & \\ 
political affiliation & 1.988$^{***}$ \\ 
 & (0.475) \\ 
 & \\ 
percentage unemployed & 16.336$^{***}$ \\ 
 & (3.852) \\ 
 & \\ 
mean distance to hospital & $-$23.773$^{***}$ \\ 
 & (2.462) \\ 
 & \\ 
median household income & $-$7.408$^{***}$ \\ 
 & (2.748) \\ 
 & \\ 
percentage AIAN & $-$33.687$^{***}$ \\ 
 & (12.145) \\ 
 & \\ 
percentage NHPI & 0.112 \\ 
 & (5.062) \\ 
 & \\ 
\hline \\[-1.8ex] 
observations  & 3,212 \\ 
R$^{2}$ & 0.552 \\ 
adjusted R$^{2}$ & 0.546 \\ 
residual std. error & 3,405.648 (df = 3170) \\ 
\hline 
\hline \\[-1.8ex] 
\textit{Note:}  & \multicolumn{1}{r}{$^{*}$p$<$0.1; $^{**}$p$<$0.05; $^{***}$p$<$0.01} \\ 
\end{tabular} 
\end{table}

\begin{table}[!htbp] \centering 
  \caption{\footnotesize{Two-way fixed-effect model for counties in the western and central United States}}
  \label{tab:two-way-fixed-effect-western-united-states} 
  \tiny
\begin{tabular}{@{\extracolsep{5pt}}lc} 
\\[-1.8ex]\hline 
\hline \\[-1.8ex] 
 & \multicolumn{1}{c}{\textit{Dependent variable: death rate per 100,000 people}} \\ 
\cline{2-2} \\
\hline \\[-1.8ex] 
deaths social proximity & 2.335$^{***}$ \\ 
 & (0.193) \\ 
 & \\ 
deaths spatial proximity & 0.011 \\ 
 & (0.355) \\ 
 & \\ 
ODR & 471.881$^{***}$ \\ 
 & (54.213) \\ 
 & \\ 
naloxone available & 3,058.806$^{***}$ \\ 
 & (610.337) \\ 
 & \\ 
buprenorphine available & 0.0002$^{***}$ \\ 
 & (0.0001) \\ 
 & \\ 
state count illicit opioid reported & $-$365.014 \\ 
 & (11,282.540) \\ 
 & \\ 
population density & 0.003$^{***}$ \\ 
 & (0.0002) \\ 
 & \\ 
mental health distress rate & 174.873$^{***}$ \\ 
 & (25.425) \\ 
 & \\ 
political affiliation & $-$0.162 \\ 
 & (0.275) \\ 
 & \\ 
percentage unemployed & 3.459 \\ 
 & (2.526) \\ 
 & \\ 
mean distance to hospital & $-$6.863$^{***}$ \\ 
 & (1.357) \\ 
 & \\ 
median household income & 8.489$^{***}$ \\ 
 & (1.786) \\ 
 & \\ 
percentage AIAN & $-$23.412$^{***}$ \\ 
 & (3.021) \\ 
 & \\ 
percentage NHPI & $-$1.874 \\ 
 & (1.157) \\ 
 & \\ 
\hline \\[-1.8ex] 
observations  & 3,004 \\ 
R$^{2}$ & 0.510 \\ 
adjusted R$^{2}$ & 0.504 \\ 
residual std. error & 1625.446 (df = 2967) \\  
\hline 
\hline \\[-1.8ex] 
\textit{Note:}  & \multicolumn{1}{r}{$^{*}$p$<$0.1; $^{**}$p$<$0.05; $^{***}$p$<$0.01} \\ 
\end{tabular} 
\end{table}

\begin{table}[!htbp] \centering 
  \caption{\footnotesize{Two-way fixed-effect model for counties in the contiguous United States}}
  \label{tab:two-way-fixed-effect-entire-united-states} 
  \tiny
\begin{tabular}{@{\extracolsep{5pt}}lc} 
\\[-1.8ex]\hline 
\hline \\[-1.8ex] 
 & \multicolumn{1}{c}{\textit{Dependent variable: death rate per 100,000 people }} \\ 
\cline{2-2} \\
\hline \\[-1.8ex] 
deaths social proximity & 4.065$^{***}$ \\ 
 & (0.317) \\ 
 & \\ 
deaths spatial proximity & 1.741$^{**}$ \\ 
 & (0.720) \\ 
 & \\ 
ODR & 141.232$^{***}$ \\ 
 & (26.094) \\ 
 & \\ 
naloxone available & 8,212.523$^{***}$ \\ 
 & (538.807) \\ 
 & \\ 
buprenorphine available & 0.001$^{***}$ \\ 
 & (0.0001) \\ 
 & \\ 
state count illicit opioid reported & $-$101.839 \\ 
 & (3,171.807) \\ 
 & \\ 
population density & $-$0.00004 \\ 
 & (0.0001) \\ 
 & \\ 
mental health distress rate & 301.861$^{***}$ \\ 
 & (24.263) \\ 
 & \\ 
political affiliation & 0.258 \\ 
 & (0.304) \\ 
 & \\ 
percentage unemployed & 16.511$^{***}$ \\ 
 & (2.750) \\ 
 & \\ 
mean distance to hospital & $-$16.874$^{***}$ \\ 
 & (2.026) \\ 
 & \\ 
median household income & 6.457$^{***}$ \\ 
 & (2.019) \\ 
 & \\ 
percentage AIAN & $-$30.523$^{***}$ \\ 
 & (4.315) \\ 
 & \\ 
percentage NHPI & 0.343 \\ 
 & (1.918) \\ 
 & \\ 

\hline \\[-1.8ex] 
observations  & 6,216 \\ 
R$^{2}$ & 0.565 \\ 
adjusted R$^{2}$ & 0.560 \\ 
residual std. error & 2,889.289 (df = 6152) \\ 
\hline 
\hline 
\hline \\[-1.8ex] 
\textit{Note:}  & \multicolumn{1}{r}{$^{*}$p$<$0.1; $^{**}$p$<$0.05; $^{***}$p$<$0.01} \\ 
\end{tabular} 
\end{table} 

\clearpage
\subsection{Two-stage least squares estimation}
\label{sec:G2sls}
Our regression analysis identified two endogenous variables, $s_{-i}$ and $d_{-i}$. Initially, we attempted to correct the correlation in error terms separately, employing network and spatial autocorrelation methods. However, these approaches do not account for correlated error terms that emerge from the simultaneous estimation of effects from socially and spatially lagged variables. To enhance the robustness of our findings, we adopt a two-stage least squares (2SLS) methodology. This approach effectively handles the correlated error terms associated with including endogenous variables in the regression model. The implementation of this model proceeds in two stages. In the first stage, we regress $s_{-i}$ and $d_{-i}$ on a matrix of instrumental variables, denoted by $Q$. This matrix is defined as $Q = (Z_n, W Z_n, A Z_n, W^2 Z_n, A^2 Z_n, W A Z_n, A W Z_n)$, where $Z_n$ is an $n \times k$ matrix. Here, $n$ represents the total number of spatial units, and $k$ signifies the number of covariates. Furthermore, $W$ and $A$ are the social and spatial weight matrices, each with dimensions $n \times n$. Our choice of instruments is based on the results of Lee et al. (2010) \cite{lee2010efficient} for higher-order lags. 

The equations for the first stage are given by:
\begin{align*}
    s_{-i} &= \omega_0 + \overline{\omega_1}^T \overline{Q}_i  + u_{s_{-i}}, \\
    d_{-i} &= \delta_0 + \overline{\delta_1}^T \overline{Q}_i + u_{d_{-i}}.
\end{align*}
were $u_{s_{-i}}$ and $u_{d_{-i}}$ represent the error term for endogenous variable $s_{-i}$ and $d_{-i}$ and $Q_{i}$ represents the set of instrumental variables for $i$-th county observation. 

In the second stage, the predicted values of $s_{-i}$ and $d_{-i}$ derived from the first stage are used to fit the final model for $y_i$. The model is defined by:
\begin{align*}
    y_i = \alpha_0 + \alpha_1 \hat{s}_{-i} + \alpha_2 \hat{d}_{-i} + \overline{\alpha_3}^T \overline{C}_i + \overline{\alpha_4}^T \overline{X}_i + \epsilon_i.
\end{align*}

\begin{figure}[!hbt]
   \centering
   \includegraphics[width=\textwidth]{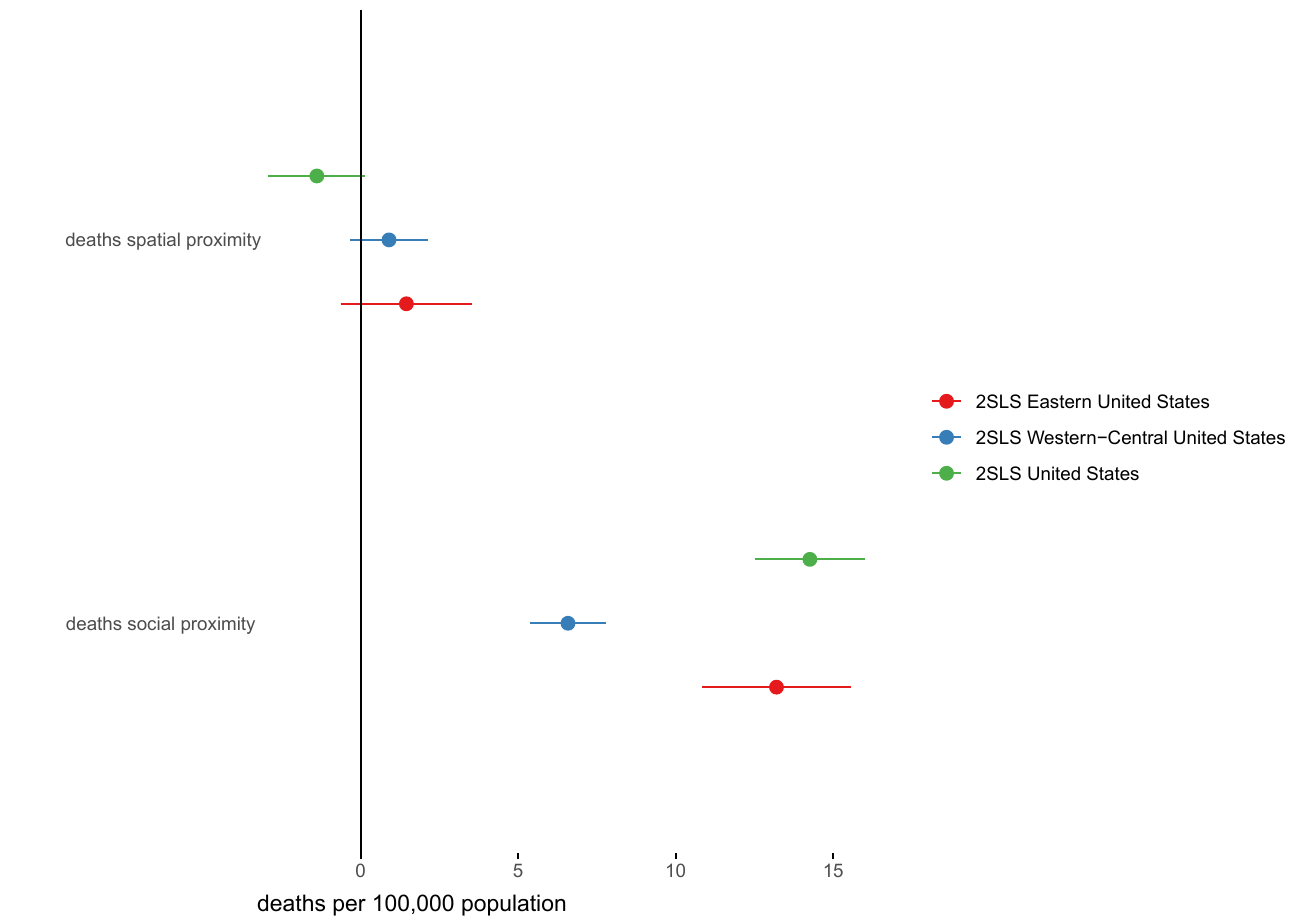}
   \caption{{This figure shows the confidence intervals for the two-stage least squares estimates, comparing the counties in the eastern, western, and entire contiguous United States. The effect sizes for $s_{-i}$ are statistically significant (p$<$ 0.001) for the three regions. The estimation results are in Supplementary Table \ref{tab:G2SLS-east-west-entire_us}.}}
   \label{fig:G2SLS-Coef-plot}
\end{figure}

Supplementary Figure \ref{fig:G2SLS-Coef-plot} shows the CIs for our 2SLS estimates of $s_{-i}$ and $d_{-i}$ effect sizes in the eastern, western-central and the entire contiguous US. The estimated effect sizes for $s_{-i}$ in all three regions are significant and positive. Supplementary Table \ref{tab:G2SLS-east-west-entire_us} provides the 2SLS estimation results and effect sizes.
\clearpage

\begin{table}[!htbp] 
\centering 
\caption{\footnotesize{This table shows the results of the two-stage least squares estimation, comparing the impact of social and spatial variables on overdose death rates in distinct geographical regions. Column (1) presents the estimates for the counties in the eastern United States , column (2) for the counties in the western and central United States, and column (3) encompasses counties in the entire contiguous United States.}} 
\label{tab:G2SLS-east-west-entire_us} 
\scriptsize
\begin{tabular}{@{\extracolsep{5pt}}lccc} 
\\[-1.8ex]\hline 
\hline \\[-1.8ex] 
 & \multicolumn{3}{c}{\textit{Dependent variable: death rate per 100,000 people}} \\ 
\cline{2-4} 
\\[-1.8ex] & (1) & (2) & (3)\\ 
\hline \\[-1.8ex] 
deaths social proximity & 13.20017$^{***}$ & 6.58148$^{***}$ & 14.25920$^{***}$ \\ 
 & (1.19966) & (0.62021) & (0.88471) \\ 
 & & & \\ 
deaths spatial proximity & 1.45265 & 0.89995 & $-$1.38995$^{*}$ \\ 
 & (1.06666) & (0.63798) & (0.78416) \\ 
 & & & \\ 
ODR & 52.22437$^{***}$ & 6.17012$^{**}$ & 38.43051$^{***}$ \\ 
 & (7.53112) & (2.57692) & (4.80499) \\ 
 & & & \\ 
naloxone available & 32.04615$^{***}$ & $-$3.78452 & 21.51028$^{***}$ \\ 
 & (11.78907) & (5.97941) & (6.64167) \\ 
 & & & \\ 
buprenorphine available & 15.08753$^{**}$ & 27.46304$^{***}$ & 23.85361$^{***}$ \\ 
 & (6.84219) & (6.06131) & (5.48752) \\ 
 & & & \\ 
state count illicit opioid reported & $-$483.76320 & 860.05570 & 1,156.91900 \\ 
 & (2,496.26400) & (3,439.58200) & (1,898.00800) \\ 
 & & & \\ 
population density & 0.00161$^{**}$ & 0.00955$^{***}$ & 0.00294$^{***}$ \\ 
 & (0.00070) & (0.00158) & (0.00058) \\ 
 & & & \\ 
mental health distress rate & $-$104.88350$^{*}$ & $-$75.99799$^{***}$ & $-$111.36910$^{***}$ \\ 
 & (57.89841) & (28.63051) & (29.82314) \\ 
 & & & \\ 
political affiliation & $-$5.34279$^{***}$ & $-$0.47798 & $-$4.48322$^{***}$ \\ 
 & (1.38088) & (1.09760) & (0.92130) \\ 
 & & & \\ 
percentage unemployed & 7.35985 & 4.65552 & 2.53671 \\ 
 & (5.72459) & (4.06029) & (3.92320) \\ 
 & & & \\ 
mean distance to hospital & $-$26.03972$^{***}$ & $-$2.17998 & $-$7.79723$^{***}$ \\ 
 & (4.18689) & (2.58041) & (2.99407) \\ 
 & & & \\ 
median household income & $-$25.27024$^{***}$ & $-$7.75322$^{*}$ & $-$12.02835$^{***}$ \\ 
 & (7.36773) & (4.00888) & (4.44148) \\ 
 & & & \\ 
percentage AIAN & 45.87866$^{***}$ & 8.02707$^{**}$ & 19.49589$^{***}$ \\ 
 & (15.17279) & (3.97502) & (4.72510) \\ 
 & & & \\ 
percentage NHPI & 5.78337 & $-$2.46150 & 1.34235 \\ 
 & (6.75925) & (4.80422) & (6.01157) \\ 
 & & & \\ 
constant & 45.35172$^{***}$ & 18.23200$^{***}$ & 33.05425$^{***}$ \\ 
 & (8.79892) & (4.39869) & (4.59876) \\ 
 & & & \\ 
\hline \\[-1.8ex] 
observations  & 1,606 & 1,502 & 3,108 \\ 
R$^{2}$ & 0.40527 & 0.30141 & 0.42091 \\ 
adjusted R$^{2}$ & 0.40003 & 0.29483 & 0.41829 \\ 
residual std. error & 19.28330 (df = 1591) & 12.20299 (df = 1487) & 16.92010 (df = 3093) \\ 
\hline 
\hline \\[-1.8ex] 
\textit{Note:} & \multicolumn{3}{r}{$^{*}$p$<$0.1; $^{**}$p$<$0.05; $^{***}$p$<$0.01} \\ 
\end{tabular} 
\end{table}

\newpage
\section{Robustness check on the spatial weight matrix}\label{sec:distance decay weight check}

\begin{figure}[!hbt]
   \centering
   \includegraphics[width=\textwidth]{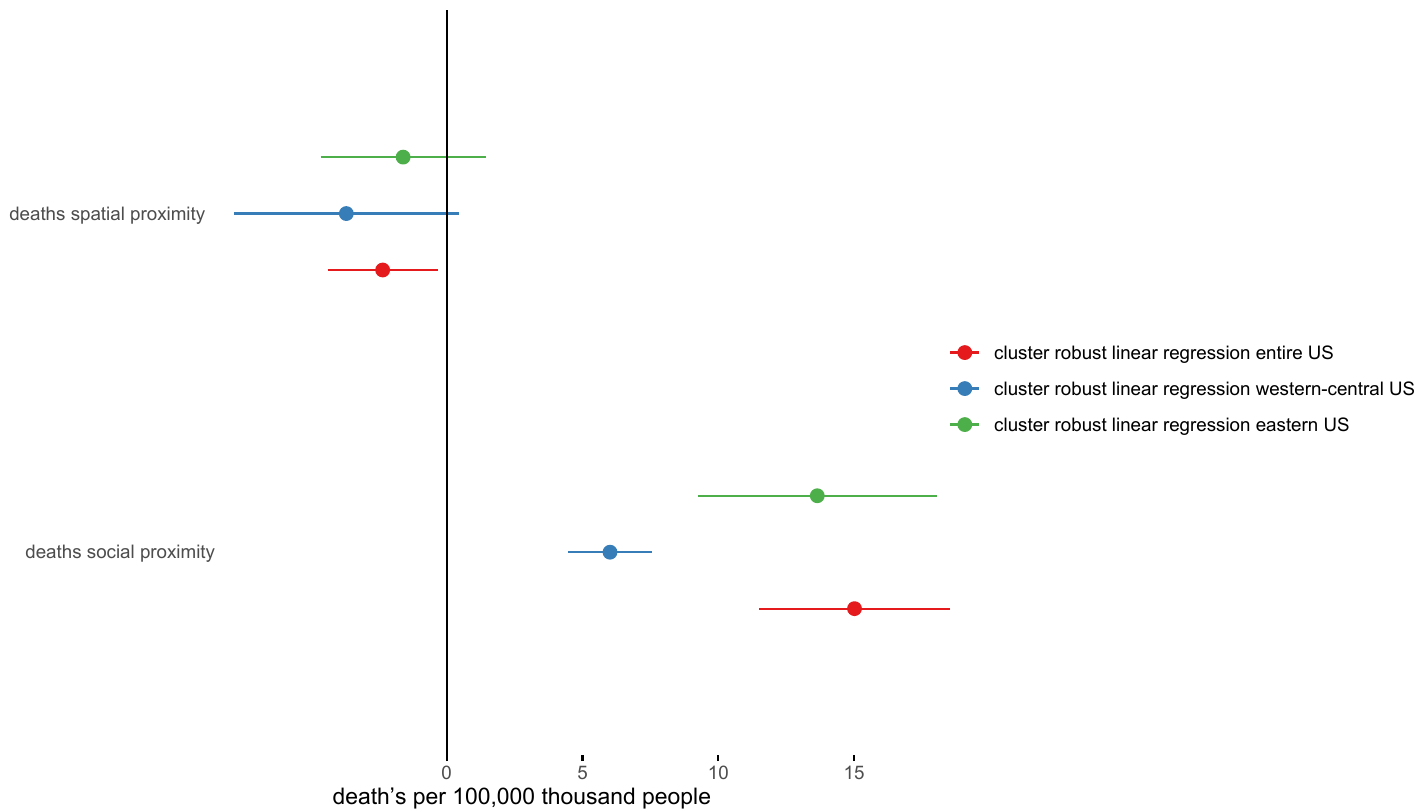}
   \caption{{This figure shows the confidence intervals for the cluster-robust linear regression with spatial proximity based on the slow-decaying distance weights, comparing the counties in the eastern, western, and entire contiguous United States. The effect sizes for $s_{-i}$ are statistically significant (p$<$ 0.001) for the three regions. The estimation results are in Supplementary Table \ref{tab:linear_regression_eastern_united_states_distance_decay_weight}, \ref{tab:linear_regression_western_united_states_distance_decay_weight} and \ref{tab:linear_regression_entire_united_states_distance_decay_weight}.}}
   \label{fig:lm-sw-Coef-plot}
\end{figure}

To enhance the robustness of our findings, we analyze the spatial adjacency matrix with the distance decay function. The major reason to do so is to ensure that the modified distance decay function reduces the rate at which adjacency weights decrease with distance. This allows us to capture the effect of farther away counties that would otherwise have been under-weighted. We define the the updated weight matrix as follows:\begin{align*}
a'_{ij} = \frac{1 + \frac{1}{d_{ij}^{^{1/10}}}}{\sum_{k \neq i} \left(1 + \frac{1}{d_{ik}^{^{1/10}}}\right)},
\end{align*} where $d_{ij}$ is the distance matrix between county $i$ and $j$. We run the cluster robust linear regression for counties in eastern US, western and central US pooled together, and counties in the contiguous US. Table \ref{tab:linear_regression_eastern_united_states_distance_decay_weight}, \ref{tab:linear_regression_western_united_states_distance_decay_weight} and \ref{tab:linear_regression_entire_united_states_distance_decay_weight} consistently show the coefficient for ``deaths in social proximity" is statistically significant and adds additional robustness to our analysis. Supplementary figure \ref{fig:lm-sw-Coef-plot}
shows the confidence interval plot highlighting the statistically significant effect of coefficient for ``deaths in social proximity". 

\begin{table}[!htbp] \centering 
  \caption{Linear regression for measuring the significance of the effect size of ``deaths in social proximity", for counties in the eastern United States with slowly decaying distance weights} 
  \label{tab:linear_regression_eastern_united_states_distance_decay_weight} 
  \scriptsize
\begin{tabular}{@{\extracolsep{5pt}}lcc} 
\\[-1.8ex]\hline 
\hline \\[-1.8ex] 
 & \multicolumn{2}{c}{\textit{Dependent variable: death rate per 100,000 people}} \\ 
\cline{2-3} 
\\[-1.8ex] & \multicolumn{2}{c}{ } \\ 
\\[-1.8ex] & \textit{OLS} & \textit{Cluster-Robust OLS} \\ 
\\[-1.8ex] & (1) & (2)\\ 
\hline \\[-1.8ex] 
deaths social proximity & 13.643$^{***}$ & 13.643$^{***}$ \\ 
 & (0.863) & (2.243) \\ 
 & & \\ 
deaths spatial proximity & $-$1.604$^{**}$ & $-$1.604 \\ 
 & (0.780) & (1.546) \\ 
 & & \\ 
ODR & 39.663$^{***}$ & 39.663 \\ 
 & (12.750) & (25.848) \\ 
 & & \\ 
naloxone available & 172.358$^{***}$ & 172.358$^{**}$ \\ 
 & (17.162) & (83.825) \\ 
 & & \\ 
buprenorphine available & 57.944$^{***}$ & 57.944 \\ 
 & (13.297) & (44.091) \\ 
 & & \\ 
state count illicit opioid reported & 10,648.380$^{***}$ & 10,648.380 \\ 
 & (3,208.401) & (10,006.390) \\ 
 & & \\ 
 population density & $-$0.0001 & $-$0.0001 \\ 
  & (0.0002) & (0.001) \\ 
  & & \\ 
 mental health distress rate & 154.389$^{**}$ & 154.389 \\ 
  & (77.943) & (152.686) \\ 
  & & \\ 
 political affiliation & $-$3.258$^{**}$ & $-$3.258 \\ 
  & (1.354) & (2.850) \\ 
  & & \\ 
 percentage unemployed & 77.127$^{***}$ & 77.127$^{**}$ \\ 
  & (9.729) & (30.586) \\ 
  & & \\ 
 mean distance to  hospital & $-$69.209$^{***}$ & $-$69.209$^{***}$ \\ 
  & (7.184) & (11.786) \\ 
  & & \\ 
 median household income & $-$12.072$^{*}$ & $-$12.072 \\ 
  & (6.896) & (14.549) \\ 
  & & \\ 
 percentage AIAN & $-$28.784 & $-$28.784 \\ 
  & (36.110) & (21.705) \\ 
  & & \\ 
 percentage NHPI & $-$19.562 & $-$19.562 \\ 
  & (12.359) & (14.146) \\ 
  & & \\ 
 Constant & $-$0.219 & $-$0.219 \\ 
  & (10.465) & (22.483) \\ 
  & & \\ 
\hline \\[-1.8ex] 
observations  & 1,606 &  \\ 
R$^{2}$ & 0.454 &  \\ 
adjusted R$^{2}$ & 0.449 &  \\ 
residual std. error & 3.210 (df = 1590) &  \\ 
F statistic & 94.369$^{***}$ (df = 14; 1590) &   \\ 
\hline 
\hline \\[-1.8ex] 
\textit{Note:}  & \multicolumn{2}{r}{$^{*}$p$<$0.1; $^{**}$p$<$0.05; $^{***}$p$<$0.01} \\ 
\end{tabular} 
\end{table}

\begin{table}[!htbp] \centering 
  \caption{\footnotesize{Linear regression for measuring the significance of the effect size of ``deaths in social proximity", for counties in the western and central United States with slowly decaying distance weights}} 
\label{tab:linear_regression_western_united_states_distance_decay_weight} 
\scriptsize
\begin{tabular}{@{\extracolsep{5pt}}lcc} 
\\[-1.8ex]\hline 
\hline \\[-1.8ex] 
 & \multicolumn{2}{c}{\textit{Dependent variable: death rate per 100,000 people}} \\ 
\cline{2-3} 
\\[-1.8ex] & \textit{OLS} & \textit{Cluster-Robust OLS} \\ 
\\[-1.8ex] & (1) & (2)\\ 
\hline \\[-1.8ex] 
 deaths social proximity & 6.013$^{***}$ & 6.013$^{***}$ \\ 
 & (0.234) & (0.781) \\ 
 & & \\ 
deaths spatial proximity & $-$3.694$^{***}$ & $-$3.694$^{*}$ \\ 
 & (0.375) & (2.108) \\ 
 & & \\ 
ODR & 14.954$^{***}$ & 14.954 \\ 
 & (3.734) & (9.451) \\ 
 & & \\ 
naloxone available & 93.746$^{***}$ & 93.746$^{**}$ \\ 
 & (10.214) & (39.755) \\ 
 & & \\ 
buprenorphine available & 46.277$^{***}$ & 46.277$^{**}$ \\ 
 & (8.315) & (19.070) \\ 
 & & \\ 
state count illicit opioid reported & $-$8,620.923$^{**}$ & $-$8,620.923 \\ 
 & (4,324.188) & (12,994.870) \\ 
 & & \\ 
population density & 0.005$^{***}$ & 0.005$^{***}$ \\ 
  & (0.0005) & (0.001) \\ 
  & & \\ 
 mental health distress rate & 92.493$^{**}$ & 92.493 \\ 
  & (37.035) & (126.872) \\ 
  & & \\ 
 political affiliation & $-$0.472 & $-$0.472 \\ 
  & (0.704) & (2.299) \\ 
  & & \\ 
 percentage unemployed & 15.921$^{***}$ & 15.921 \\ 
  & (5.979) & (17.045) \\ 
  & & \\ 
 mean distance to hospital & $-$9.119$^{***}$ & $-$9.119 \\ 
  & (3.513) & (10.910) \\ 
  & & \\ 
 median household income & 6.268$^{*}$ & 6.268 \\ 
  & (3.249) & (10.881) \\ 
  & & \\ 
 percentage AIAN & $-$12.689$^{*}$ & $-$12.689 \\ 
  & (6.560) & (11.559) \\ 
  & & \\ 
 percentage NHPI & 12.089$^{***}$ & 12.089 \\ 
  & (3.693) & (10.008) \\ 
  & & \\ 
 Constant & $-$10.105$^{**}$ & $-$10.105 \\ 
  & (5.011) & (20.221) \\ 
  & & \\ 
\hline \\[-1.8ex] 
observations  & 1,502 &  \\ 
R$^{2}$ & 0.492 &  \\ 
adjusted R$^{2}$ & 0.487 &  \\ 
residual std. error & 1.004 (df = 1486) &  \\ 
F statistic & 102.823$^{***}$ (df = 14; 1486) &  \\ 
\hline 
\hline \\[-1.8ex] 
\textit{Note:}  & \multicolumn{2}{r}{$^{*}$p$<$0.1; $^{**}$p$<$0.05; $^{***}$p$<$0.01} \\ 
\end{tabular} 
\end{table}

\begin{table}[!htbp] \centering 
  \caption{\footnotesize{Linear regression for measuring the significance of the effect size of ``deaths in social proximity", for counties in the contiguous United States with slowly decaying distance weights}}  \label{tab:linear_regression_entire_united_states_distance_decay_weight} 
  \scriptsize
\begin{tabular}{@{\extracolsep{5pt}}lcc} 
\\[-1.8ex]\hline 
\hline \\[-1.8ex] 
 & \multicolumn{2}{c}{\textit{Dependent variable: death rate per 100,000 people}} \\ 
\cline{2-3} 
\\[-1.8ex] & death rates &   \\ 
\\[-1.8ex] & \textit{OLS} & \textit{Cluster-Robust OLS} \\  
\\[-1.8ex] & (1) & (2)\\ 
\hline \\[-1.8ex] 
deaths social proximity & 15.013$^{***}$ & 15.013$^{***}$ \\ 
 & (0.622) & (1.787) \\ 
 & & \\ 
deaths spatial proximity & $-$2.357$^{***}$ & $-$2.357$^{**}$ \\ 
 & (0.582) & (1.035) \\ 
 & & \\ 
ODR & 48.719$^{***}$ & 48.719$^{**}$ \\ 
 & (8.169) & (19.212) \\ 
 & & \\ 
naloxone available & 150.050$^{***}$ & 150.050$^{***}$ \\ 
 & (11.250) & (56.853) \\ 
 & & \\ 
buprenorphine available & 60.425$^{***}$ & 60.425 \\ 
 & (10.014) & (42.605) \\ 
 & & \\ 
state count illicit opioid reported & 13,248.850$^{***}$ & 13,248.850 \\ 
 & (2,365.555) & (9,362.013) \\ 
 & & \\ 
population density & 0.001$^{***}$ & 0.001 \\ 
 & (0.0002) & (0.001) \\ 
 & & \\ 
mental health distress rate & 69.993 & 69.993 \\ 
 & (45.879) & (112.128) \\ 
 & & \\ 
political affiliation & $-$3.194$^{***}$ & $-$3.194$^{*}$ \\ 
 & (0.828) & (1.884) \\ 
 & & \\ 
percentage unemployed & 42.330$^{***}$ & 42.330$^{*}$ \\ 
 & (6.718) & (23.953) \\ 
 & & \\ 
mean distance to hospital & $-$49.510$^{***}$ & $-$49.510$^{***}$ \\ 
 & (5.545) & (13.203) \\ 
 & & \\ 
median household income & $-$12.136$^{***}$ & $-$12.136 \\ 
 & (4.271) & (9.640) \\ 
 & & \\ 
percentage AIAN & $-$26.900$^{**}$ & $-$26.900$^{*}$ \\ 
 & (11.020) & (14.641) \\ 
 & & \\ 
percentage NHPI & $-$18.952$^{***}$ & $-$18.952$^{*}$ \\ 
 & (5.748) & (10.007) \\ 
 & & \\ 
constant & 1.248 & 1.248 \\ 
 & (6.158) & (15.344) \\ 
 & & \\ 
\hline \\[-1.8ex] 
observations  & 3,108 &  \\ 
R$^{2}$ & 0.504 &  \\ 
adjusted R$^{2}$ & 0.502 &  \\ 
residual std. error & 1.895 (df = 3090) &  \\ 
F statistic & 224.521$^{***}$ (df = 14; 3090) &   \\ 
\hline 
\hline \\[-1.8ex] 
\textit{Note:}  & \multicolumn{2}{r}{$^{*}$p$<$0.1; $^{**}$p$<$0.05; $^{***}$p$<$0.01} \\ 
\end{tabular} 
\end{table} 

\end{document}